\documentclass[12pt,subeqn]{article}

\usepackage{amsmath}

\renewcommand{\theequation}{\arabic{equation}}
\newcommand{\tr}{\operatorname{Tr}}
\newcommand{\diag}{\operatorname{diag}}
\newcommand{\EQ}{\begin{equation}}
\newcommand{\EN}{\end{equation}}

\newcommand{\ket}[1]{\left|#1\right\rangle}      % Ket-Zustand
\newcommand{\bear}{\begin{eqnarray}}
\newcommand{\ear}{\end{eqnarray}}
\newcommand{\bt} { \begin{tabular} }
\newcommand{\et}{ \end{tabular} }
\newcommand{\bc} { \begin{center} }
\newcommand{\ec}{ \end{center} }

\newcommand{\btb} { \begin{table} }
\newcommand{\etb}{ \end{table} }
\begin{document}

\topmargin 0pt
\oddsidemargin 5mm
\newcommand{\NP}[1]{Nucl.\ Phys.\ {\bf #1}}
\newcommand{\PL}[1]{Phys.\ Lett.\ {\bf #1}}
\newcommand{\NC}[1]{Nuovo Cimento {\bf #1}}
\newcommand{\CMP}[1]{Comm.\ Math.\ Phys.\ {\bf #1}}
\newcommand{\PR}[1]{Phys.\ Rev.\ {\bf #1}}
\newcommand{\PRL}[1]{Phys.\ Rev.\ Lett.\ {\bf #1}}
\newcommand{\MPL}[1]{Mod.\ Phys.\ Lett.\ {\bf #1}}
\newcommand{\JETP}[1]{Sov.\ Phys.\ JETP {\bf #1}}
\newcommand{\TMP}[1]{Teor.\ Mat.\ Fiz.\ {\bf #1}}

\renewcommand{\thefootnote}{\fnsymbol{footnote}}

\newpage
\setcounter{page}{0}
\begin{titlepage}
\begin{flushright}
UFSCARF-TH-04-17
\end{flushright}
\vspace{0.5cm}
\begin{center}
{\large Mixed integrable $SU(N)$ vertex model with arbitrary twists}\\
\vspace{1cm}
{\large G.A.P. Ribeiro and M.J. Martins} \\
\vspace{1cm}
{\em Universidade Federal de S\~ao Carlos\\
Departamento de F\'{\i}sica \\
C.P. 676, 13565-905~~S\~ao Carlos(SP), Brasil}\\
\end{center}
\vspace{0.5cm}

\begin{abstract}
We consider the quantum inverse scattering method for 
several mixed integrable models based on the rational $SU(N)$ $R$-matrix with general toroidal boundary conditions. This includes systems whose Hilbert spaces are 
invariant by the discrete representations of the group $SU(2)$ and 
the non-compact group $SU(1,1)$ as well as the conjugate representation of the $SU(N)$ symmetry. Introducing 
certain transformations on the quantum spaces we are able to solve 
generalized impurity problems including those related to singular matrices.
\end{abstract}

\vspace{.15cm}
\centerline{PACS numbers:  05.50+q, 02.30.IK}
\vspace{.1cm}
\centerline{Keywords: Algebraic Bethe Ansatz, Mixed Lattice Models, Boundary Conditions}
\vspace{.15cm}
\centerline{June 2004}

\end{titlepage}

%\tableofcontents

\pagestyle{empty}

\newpage

\pagestyle{plain}
\pagenumbering{arabic}

\renewcommand{\thefootnote}{\arabic{footnote}}

\section{Introduction}

The quantum version of the inverse scattering method is nowadays accepted as the foundation of the algebraic theory of integrable models of quantum field theory and classical statistical mechanics in $(1+1)$ dimensions \cite{FA,KO}. This approach has made possible not only the discovery of new important models such as the Heisenberg chain with arbitrary spin \cite{KUL,BATA} but also precipitated the notion of quantum group symmetry \cite{QG1,QG2}.

An essential ingredient of the quantum inverse scattering method is frequently denominated the Yang-Baxter algebra whose generators can be thought as the elements of square $N\times N$ matrix ${\cal T_{A}}(\lambda)$. The symbol $\cal A$ emphasizes such $N$-dimensional auxiliary space and $\lambda$ denotes the spectral parameter. This associative algebra is then generated by the following quadratic relations
\EQ
{\check{R}}(\lambda- \mu) {\cal T}_{{\cal A}}(\lambda) 
\otimes {\cal T}_{{\cal A}}(\mu) = {\cal T}_{{\cal A}}(\mu) \otimes {\cal T}_{{\cal A}}(\lambda) {\check{R}}(\lambda- \mu),
\label{fundrel}
\EN
where ${\check{R}}(\lambda)$ is a $N^{2}\times N^{2}$ matrix whose elements are complex numbers. Here we are assuming that this $R$-matrix is additive with respect the spectral parameters.

The $R$-matrix ${\check{R}}(\lambda)$ is required to satisfy a 
sufficient condition, that guarantees the associativity of the algebra
(\ref{fundrel}), known as the Yang-Baxter equation
\EQ
{\check{R}}_{12}(\lambda){\check{R}}_{23}(\lambda+\mu){\check{R}}_{12}(\mu)={\check{R}}_{23}(\mu){\check{R}}_{12}(\lambda+\mu){\check{R}}_{23}(\lambda),
\label{yangbaxter}
\EN
where ${\check{R}}_{ab}(\lambda)$ denotes the action of the 
$R$-matrix on the tensor product space ${\cal A}_{a}\otimes {\cal A}_{b}$.

It is not difficult to see that the trace of ${\cal T}_{\cal A}(\lambda)$ over the auxiliary space
\EQ
T(\lambda)=\tr_{\cal A}\left[ {\cal T}_{\cal A}(\lambda) \right],
\label{transfermatrix}
\EN
gives origin to a commutative family of operators, i.e. 
$\left[ T(\lambda), T(\mu) \right]=0$ for arbitrary values of $\lambda$ and $\mu$. In this sense, $T(\lambda)$ can be regarded as the generating function of quantum integrals of motion.

One possible representation of ${\cal T}_{\cal A}(\lambda)$ is provided by numeric $N\times N$ matrix ${\cal G}_{\cal A}$ whose entries do not depend on $\lambda$ and that satisfies the relation
\EQ
\left[ {\check{R}}(\lambda), {\cal G}_{\cal A} \otimes {\cal G}_{\cal A}\right]=0,
\label{symmetry}
\EN
and therefore being direct related to the underlying symmetries of ${\check{R}}(\lambda)$.

General representations of (\ref{fundrel}) for a given $R$-matrix depend on the spectral parameter $\lambda$ and are often  called $\cal L$-operators ${\cal L}_{{\cal A}j}(\lambda)$. They are viewed as 
matrices on the auxiliary space $\cal A$ whose elements are operators on 
another space $V_{j}$ known as quantum space. The simplest one occurs when the auxiliary space $\cal A$ and the quantum space $V_{j}$ are isomorphic, since in this case Eq.(\ref{yangbaxter}) becomes equivalent to Eq.(\ref{fundrel}) provided we set
\EQ
{\cal L}_{{\cal A}j}(\lambda)=P_{{\cal A}j}{\check{R}}(\lambda),
\label{VisoA}
\EN
where $P_{{\cal A}j}$ is the exchange operator on the space ${\cal A}\otimes V_{j}$.

An extremely important property of the Yang-Baxter algebra is that the tensor product of two representations is still a representation of this algebra. For example, let $\displaystyle {\cal L}_{{\cal A}j} \in \prod_{j=1}^{L}\otimes V_{j}$ for $L$ distinct $\cal L$-operators which satisfy the Yang-Baxter algebra with the same $R$-matrix. Then the following ordered tensor product of $\cal L$-operators, denominated monodromy matrix \cite{FA,KO}
\EQ
{\cal T}_{\cal A}(\lambda)= {\cal G}_{\cal A} {\cal L}_{{\cal A} L}(\lambda){\cal L}_{{\cal A} L-1}(\lambda)\dots {\cal L}_{{\cal A} 1}(\lambda),
\label{generalmonodromy}
\EN
is also a representation of the same Yang-Baxter algebra.

In the context of classical vertex models of statistical mechanics ${\cal L}_{{\cal A}j}(\lambda)$ represents the possible Boltzmann weights at the local site $j$ of a square lattice of size $L$, ${\cal G_{A}}$ play the role of possible toroidal boundary conditions compatible with integrability \cite{DE,BA1} and the operator $T(\lambda)$ turns out to be the corresponding row-to-row transfer matrix. The purpose of this paper is to investigate the eigenvalue problem for the transfer matrix $T(\lambda)$ of such mixed vertex models whose underlying $R$-matrix is the simplest rational solution of the Yang-Baxter equation invariant by the fundamental representation of $SU(N)$, namely \cite{SUN,SUN1}
\EQ
{\check{R}}_{ab}(\lambda)=\eta \sum_{\alpha=1}^{N} \hat{e}_{\alpha \alpha}^{(a)} \otimes \hat{e}_{\alpha \alpha}^{(b)} + \lambda \sum_{\alpha,\beta=1}^{N} \hat{e}_{\alpha \beta}^{(a)} \otimes \hat{e}_{\beta \alpha}^{(b)},
\label{R-matrix}
\EN
where $\hat{e}_{\alpha \beta}^{(a)}$ are the $N\times N$ Weyl matrices acting on the space ${\cal A}_{a}$  and $\eta$ is the so-called quasi-classical parameter. It turns out that the admissible boundary representations $\cal G_{A}$ for this $R$-matrix are arbitrary $N\times N$ matrices that will be represented by
\EQ
{\cal G}_{\cal A}=\sum_{\alpha,\beta=1}^{N} g_{\alpha \beta} \hat{e}_{\alpha \beta}^{({\cal A})}.
\EN

The purpose of this paper is twofold. 
On the one hand, we extend our recent efforts \cite{RMG} in solving 
fundamental $SU(N)$ vertex models with rather general non-diagonal 
toroidal boundary conditions to other interesting group invariance  of the quantum space 
that are not isomorphic to the fundamental 
representation of the $SU(N)$ symmetry. Secondly, we explore the consequences 
of the existence of transformations on the quantum space to clarify 
possible relations between the eigenvectors of $T(\lambda)$ with diagonal and non-diagonal boundary conditions. 
This fact makes possible the solution of generalized mixed vertex models 
that combine both boundary and spectral 
dependent representations at any site of the chain.

This work has been organized as follows. In section \ref{su2} we 
present the algebraic Bethe ansatz solution of $SU(2)$ mixed vertex models when their 
Hilbert spaces are invariant by higher spin representations of $SU(2)$ and the discrete $D^{+}(k)$ 
representation of the non-compact group $SU(1,1)$. In section \ref{SUNsection} 
we discuss 
similar analysis for the isotropic vertex model 
that mixes the fundamental and the 
conjugate representations of $SU(N)$. In section \ref{QSTsection} 
we show that suitable transformations on the quantum 
space are essential for the solution of general classes of mixed vertex models. This includes 
those whose boundary representations are singular as well as the integrable spin-$S$ 
Heisenberg model with the most arbitrary non-diagonal boundary conditions. The appendix A is
reserved for the study of the completeness of the Hilbert space of some of the mixed 
vertex models mentioned
above with 
$L=2$. In appendix B we present the explicit expressions of 
the quantum transformations for all the mixed systems described in this paper.

\section{Mixed $SU(2)$ vertex models}\label{su2}

The purpose of this section is to solve the eigenvalues 
problem for the transfer matrix (\ref{transfermatrix}) of mixed 
vertex models whose underlying $R$-matrix (\ref{R-matrix}) is $SU(2)$ invariant. One 
natural way of producing such mixed vertex model is 
by choosing ${\cal L}$-operators intertwining between general 
representations of $SU(2)$. The other manner is to look 
for realizations of the Yang-Baxter algebra (\ref{fundrel}) 
in terms of different algebraic structures for the quantum space $V_{j}$, 
even those based on non-compact groups. In what follows we shall explore these both possibilities.

\subsection{The higher spin realization}\label{higherspinsection}

The approach of using arbitrary representations of $SU(2)$ to built up mixed vertex models goes probably back to the work by Andrei and Johannesson \cite{AJ} who studied the spin-$1/2$ Heisenberg model in the presence of an impurity of spin-$S$. Latter on, de Vega and Woyanorovich \cite{DW} have used similar idea to construct integrable Heisenberg chains with
alternating spins resembling ferrimagnetic models. However, those works have assumed explicitly periodic 
boundary conditions and therefore not dealt with the most general forms of 
integrable impurities that are going to  be considered here.

The higher spin realization of (\ref{fundrel}) is built from the local $\cal L$-operators \cite{KUL,BATA}
\EQ
{\cal L}_{{\cal A}j}^{(\frac{1}{2},S)}(\lambda)=\left(
\begin{array}{cc}
       \left[\lambda+\frac{\eta}{2}\right] +\eta S_{j}^{z} & \eta S_{j}^{-} \\
       \eta S_{j}^{+} & \left[\lambda +\frac{\eta}{2}\right] -\eta S_{j}^{z} \end{array}\right),
\label{Lhigherspin}
\EN
which act in the tensor product of the auxiliary space of spin-$1/2$ and the space $V_{j}$ carrying the spin-$S$ representation at the $j$-th site. The construction of the transfer matrix is standard as described in the introduction and it is given by
\EQ
T_{1/2}^{(S_{1},\dots,S_{L})}(\lambda)=\tr_{\cal A}\left[ {\cal G}_{\cal A} {\cal L}_{{\cal A} L}^{(\frac{1}{2}, S_{L})}(\lambda)\dots {\cal L}_{{\cal A} 1}^{(\frac{1}{2}, S_{1})}(\lambda) \right].
\label{transferT12S}
\EN

In order to diagonalize this operator by the quantum inverse scattering method it is desirable the existence of local vacuum vectors $\ket{0}_{j}$ such that the action of the monodromy matrix ${\cal T}_{\cal A}(\lambda)$ on the global state \EQ
\ket{0}=\prod_{j=1}^{L} \otimes \ket{0}_{j},
\label{kettensor}
\EN gives as result a triangular matrix.

Since the boundary matrix $\cal G_{A}$ is arbitrary, the tensor product of the standard highest spin-$S_{j}$ states does not  work as an appropriate global reference state. Following our previous work \cite{RMG}, suitable local states can be found by introducing a set of Baxter's gauge transformations $M_{j}$ \cite{BA} on the spin-$1/2$ auxiliary space such that
\EQ
\widetilde{{\cal L}}_{{\cal A} j}^{(\frac{1}{2}, S_{j})}(\lambda)=M_{j+1}^{-1} {\cal L}_{{\cal A} j}^{(\frac{1}{2}, S_{j})}(\lambda) M_{j},
\label{gaugetransf}
\EN
where $M_{j}$ is an invertible $2\times 2$ matrix whose entries are denoted by
\EQ
M_{j}=\left(\begin{array}{cc}
        x_{j} & r_{j} \\
        y_{j} & s_{j} \\
        \end{array}\right).
\label{mj}
\EN

In terms of these new $\widetilde{\cal L}$-operators transfer matrix (\ref{transferT12S}) becomes
\EQ
T_{1/2}^{(S_{1},\dots,S_{L})}(\lambda)=\tr_{\cal A}\left[ M_{1}^{-1}{\cal G}_{\cal A}M_{L+1} \widetilde{\cal T}_{\cal A}(\lambda) \right],
\label{transferT12S-2}
\EN
where the $\widetilde{\cal T}_{\cal A}(\lambda)$ is the transformed monodromy matrix $\widetilde{\cal T}_{\cal A}(\lambda)=\widetilde{\cal L}_{{\cal A} L}^{(\frac{1}{2}, S_{L})}(\lambda)\dots \widetilde{\cal L}_{{\cal A} 1}^{(\frac{1}{2}, S_{1})}(\lambda)$.

We now search for gauge transformations $M_{j}$ in such way that 
$\widetilde{\cal L}_{{\cal A} j}^{(\frac{1}{2}, S_{j})}(\lambda)$ 
is annihilated by say its lower left element for general values of the 
spectral parameter $\lambda$. This property 
turns out to be equivalent to the following condition for the ratio $p_{j}=\frac{x_{j}}{y_{j}}$
\EQ
\left\{ \left[\lambda+\frac{\eta}{2} \right] (p_{j+1}-p_{j}) - \eta S_{j}^{z} (p_{j+1}+p_{j}) +\eta S_{j}^{+} p_{j+1} p_{j} - \eta S_{j}^{-} \right\} \ket{0}_{j}^{(S_{j})}=0.
\label{Cannihilate}
\EN

This problem can be solved by expanding the vector $\ket{0}_{j}^{(S_{j})}$ in terms of simultaneous eigenkets  $\ket{S_{j},m_{j}}$ of $\vec{S}_{j}^{2}$ and $S_{j}^{z}$,
\EQ
\ket{0}_{j}^{(S_{j})}=\sum_{m_{j}=-S_{j}}^{S_{j}} C(S_{j},m_{j}) \ket{S_{j},m_{j}}.
\EN

Substituting this ansatz in Eq.(\ref{Cannihilate}) and by taking into account the well known action of the angular momenta ladder operators $S_{j}^{\pm}$ on the basis $\ket{S_{j},m_{j}}$ we find $(2S+1)$ restrictions for the coefficients $C(S_{j},m_{j})$ given by
\bear
\sum_{m_{j}=-S_{j}}^{S_{j}} [ -(p_{j+1} + p_{j}) m_{j} C(S_{j},m_{j}) + p_{j}p_{j+1} C(S_{j},m_{j}-1) \sqrt{(S_{j}+m_{j})(S_{j}-m_{j}+1)} \nonumber \\
 (\lambda+\frac{\eta}{2})(p_{j+1} - p_{j}) C(S_{j},m_{j}) - C(S_{j},m_{j}+1) \sqrt{(S_{j}-m_{j})(S_{j}+m_{j}+1)} ] \ket{S_{j},m_{j}}=0.
\label{recurrence}
\ear

A necessary condition for the solution of these 
equations such that $\ket{0}_{j}^{(S_{j})}$ is independent of the spectral 
parameter occurs when $p_{j+1}=p_{j}$ and therefore the ratio $\frac{x_{j}}{y_{j}}$ 
for $j=1,\dots,L$ is keep fixed. If we substitute 
this condition in Eq.(\ref{recurrence}) we end up with recurrence relations for 
the coefficients $C(S_{j},m_{j})$ which can be solved in terms of an arbitrary
normalization $C(S_j,S_j)$ constant. The final result for the 
local vacuum $\ket{0}_{j}^{(S_{j})}$ can then be written as follows
\EQ
\ket{0}_{j}^{(S_{j})}=\sum_{m_{j}=-S_{j}}^{S_{j}} \sqrt{\frac{(2S_{j})!}{(S_{j}+m_{j})! (S_{j}-m_{j})!}} p_{j}^{m_{j}-S_{j}} C(S_{j},S_{j}) \ket{S_{j}, m_{j}},
\label{reference}
\EN
and the action of the operator $\widetilde{\cal L}_{{\cal A}j}^{(S_{j})}(\lambda)$ in this state is given by
\EQ
\widetilde{\cal L}_{{\cal A}j}^{(S_{j})}(\lambda)\ket{0}_{j}^{(S_{j})}=\left(\begin{array}{cc}
          \left( \lambda+\frac{\eta}{2}+ \eta S_{j} \right) \frac{y_{j}}{y_{j+1}} \ket{0}_{j}^{(S_{j})}  &  \# \\
          0  &  \left( \lambda+\frac{\eta}{2} - \eta S_{j} \right) \frac{y_{j+1}}{y_{j}} \frac{\det{(M_{j})}}{\det{(M_{j+1})}} \ket{0}_{j}^{(S_{j})}\end{array}\right),
\label{actiontilde}
\EN
where the symbol $\#$ denotes non-null states.

The constant ratio $p_{j}$ is then selected out by imposing that the transformed boundary $M_{1}^{-1}{\cal G_{A}}M_{L+1}$ matrix becomes a diagonal matrix. This restriction lead us to two possible values for $p_{j}=p^{(\pm)}$, namely
\EQ
p^{(\pm)}=\frac{(g_{11}-g_{22}) \pm \sqrt{(g_{11}-g_{22})^2 + 4g_{12}g_{21}}}{2g_{21}},
\label{ppm}
\EN
and consequently with the help of Eqs.(\ref{kettensor},\ref{reference}) to two choices for the global reference state $\ket{0}^{(\pm)}$.

We now have the basic ingredients to turn to the algebraic Bethe ansatz solution 
of the eigenvalue problem associated to $T_{1/2}^{(S_{1},\dots,S_{L})}(\lambda)$. The first step is to seek for a convenient representation for the monodromy matrix $\widetilde{\cal T_{A}}(\lambda)$. Previous experience in dealing with the quantum inverse scattering method for two-dimensional auxiliary spaces \cite{FA,KO} suggest us to adopt the form
\EQ
\widetilde{\cal T_{A}}(\lambda)=\left( \begin{array}{cc}
            \widetilde{A}(\lambda) & \widetilde{B}(\lambda) \\
            \widetilde{C}(\lambda) & \widetilde{D}(\lambda)
            \end{array}\right).
\EN

The property (\ref{actiontilde}) help us to 
identify the elements of the transformed monodromy 
matrix that acts as particle  creation and annihilation 
fields over the pseudo-vacuum $\ket{0}^{(\pm)}$. We see that $\widetilde{B}(\lambda)$ 
play the role of creation field and $\widetilde{C}(\lambda)$ is an annihilation operator
thanks to the property $\widetilde{C}(\lambda)\ket{0}^{(\pm)}=0$. On the other hand, the diagonal fields
satisfy the following important relations
\bear
\widetilde{A}(\lambda)\ket{0}^{(\pm)}&=&\prod_{j=1}^{L} \left(\lambda +\frac{\eta}{2}+\eta S_{j} \right) \frac{y_{1}}{y_{L+1}} \ket{0}^{(\pm)}, \\
\widetilde{D}(\lambda)\ket{0}^{(\pm)}&=&\prod_{j=1}^{L} \left(\lambda +\frac{\eta}{2}-\eta S_{j} \right) \frac{y_{L+1}}{y_{1}} \frac{\det{(M_{1})}}{\det{(M_{L+1})}} \ket{0}^{(\pm)}.
\ear

From these properties and the help of Eq.(\ref{ppm}) it is not 
difficult to see that the states $\ket{0}^{(\pm)}$ are themselves an eigenstate of 
$T_{1/2}^{(S_{1},\dots,S_{L})}(\lambda)$ with 
eigenvalue $\Lambda_{0\pm}^{(S_{1},\dots,S_{L})}(\lambda)$ given by
\EQ
\Lambda_{0_{\pm}}^{(S_{1},\dots,S_{L})}(\lambda)= g_{1/2}^{(\pm)} \prod_{j=1}^{L} \left(\lambda +\frac{\eta}{2}+\eta S_{j} \right)
+g_{1/2}^{(\mp)} \prod_{j=1}^{L} \left(\lambda +\frac{\eta}{2}-\eta S_{j} \right),
\EN
where the phase factors $g_{1/2}^{(\pm)}$ are the eigenvalues of the matrix $\cal G_{A}$
\EQ
g_{1/2}^{(\pm)}=\frac{(g_{11}+g_{22}) \pm \sqrt{(g_{11}-g_{22})^2 + 4g_{12}g_{21}}}{2}.
\label{gpmcont}
\EN

This result strongly suggest us that the arbitrary eigenvectors $\ket{\phi}^{(\pm)}$ of 
$T_{1/2}^{(S_{1},\dots,S_{L})}(\lambda)$ can be put in the form
\EQ
\ket{\phi}^{(\pm)}=\prod_{j=1}^{n_{\pm}} \widetilde{B}(\lambda_{j}^{(\pm)})\ket{0}^{(\pm)}.
\EN

The symmetry $\left[ {\check{R}}(\lambda), M_{j}\otimes M_{j}\right]=0$ implies that the gauge transformed monodromy matrix $\widetilde{\cal T_{A}}(\lambda)$ satisfies the Yang-Baxter algebra (\ref{fundrel}) with the same $R$-matrix as the original monodromy matrix ${\cal T_{A}}(\lambda)$. As consequence of that, the matrix elements of $\widetilde{\cal T_{A}}(\lambda)$ satisfies the same set of commutation rules of the periodic six-vertex model \cite{FA,KO} 
and from now on the main steps in the eigenvalue solution of 
$T_{1/2}^{(S_{1},\dots,S_{L})}(\lambda)$ become very similar to that 
of this well known vertex system. Considering that these details have appeared in many 
different places in the literature, see for instance ref. \cite{FA,KO}, here we shall 
present only the final results.
By imposing that $\ket{\phi}^{(\pm)}$ are eigenstates of 
$T_{1/2}^{(S_{1},\dots,S_{L})}(\lambda)$ 
and performing the convenient displacement
$\lambda_{i}^{(\pm)} \rightarrow \lambda_{i}^{(\pm)}-\frac{\eta}{2}$
we find that the corresponding eigenvalues 
$\Lambda_{n_{\pm}}^{(S_{1},\dots,S_{L})}(\lambda)$ are given by the expression
\begin{eqnarray}
\Lambda_{n_{\pm}}^{(S_{1},\dots,S_{L})}(\lambda) &= & 
g_{1/2}^{(\pm)} \prod_{j=1}^{L} \left(\lambda +\frac{\eta}{2}+\eta S_{j} \right) 
\prod_{i=1}^{n_{\pm}}\frac{\lambda_{i}^{(\pm)}-\lambda+\frac{\eta}{2}}{\lambda_{i}^{(\pm)}-\lambda -\frac{\eta}{2}}
\nonumber \\
&& +g_{1/2}^{(\mp)} \prod_{j=1}^{L} \left(\lambda +\frac{\eta}{2}-\eta S_{j} \right) \prod_{i=1}^{n_{\pm}}\frac{\lambda-\lambda_{i}^{(\pm)}+\frac{3\eta}{2}}{\lambda-\lambda_{i}^{(\pm)}+\frac{\eta}{2}},
\label{eingmeioS}
\end{eqnarray}
where the rapidities $\lambda_{i}^{(\pm)}$ satisfy the following system of transcendental equations
\EQ
\prod_{j=1}^{L} \left(\frac{\lambda_{i}^{(\pm)}+\eta S_{j}}{\lambda_{i}^{(\pm)}-\eta S_{j}}\right)=
\frac{g_{1/2}^{(\mp)}}{g_{1/2}^{(\pm)}}
\prod_{\stackrel{l=1}{l \neq i }}^{n_{\pm}} \frac{\lambda_{i}^{(\pm)}-\lambda_{l}^{(\pm)}+\eta }{\lambda_{i}^{(\pm)}-\lambda_{l}^{(\pm)}-\eta}.
\label{betheansatz}
\EN

We would like to close this section with the following comments. First we emphasize that the integers $n_{\pm}$ 
play the role of particle numbers sectors satisfying the constraint 
$n_{\pm} \leq \displaystyle 2 \sum_{i=1}^{L} S_i $. In appendix A, we illustrate this fact 
by presenting the details of a study of the
completeness of the eigenspectrum (\ref{eingmeioS},\ref{betheansatz}) for $L=2$.
Our final observation concerns with physically interesting spin chains that 
commutes with the transfer matrix (\ref{transferT12S}). Though the integrability does not 
depend on how we distribute the $\cal L$-operators ${\cal L}_{{\cal A}j}^{(\frac{1}{2},S_{j})}(\lambda)$, 
the construction of local conserved charges commuting with $T_{1/2}^{(S_{1},\dots,S_{L})}(\lambda)$ does. 
One interesting case is when we have only one impurity $\cal L$-operator ${\cal  L}_{{\cal A}L}^{(\frac{1}{2},S)}(\lambda)$ sitting at the end of the chain of $L-1$ spin-$1/2$ ${\cal L}_{{\cal A}j}^{(\frac{1}{2},\frac{1}{2})}(\lambda)$ Boltzmann weights. Because ${\cal L}_{{\cal A}j}^{(\frac{1}{2},\frac{1}{2})}(\lambda)$ 
is proportional to the exchange operator $P_{{\cal A}j}$ at the point $\lambda=0$, we can 
produce local charges by expanding $\ln{\left[T_{1/2}^{(\frac{1}{2},\dots,\frac{1}{2},S)}(\lambda)\right]}$
around this point. The second term in this expansion is the associated Hamiltonian and its general expression reads
\EQ
{\cal H}=\frac{1}{\eta} \sum_{i=1}^{L-2} \left({\cal L}_{i,i+1}^{(\frac{1}{2},\frac{1}{2})}(0)\right)^{-1}+ \left({\cal L}_{L-1,L}^{(\frac{1}{2},S)}(0)\right)^{-1}+\frac{1}{\eta} \left({\cal L}_{L-1,L}^{(\frac{1}{2},S)}(0)\right)^{-1} {\cal G}_{L-1}^{-1} \left({\cal L}_{L-1,1}^{(\frac{1}{2},\frac{1}{2})}(0)\right)^{-1}  {\cal G}_{L-1} {\cal L}_{L-1,L}^{(\frac{1}{2},S)}(0),
\EN
where we have  implicitly assumed that the boundary matrix 
$\cal G_{A}$ is non singular. 

By substituting the expression for the weights 
${\cal L}_{{\cal{A}}j}^{(\frac{1}{2},S_j)}(\lambda)$ in the above equation 
and after several manipulations we find that
${\cal H}$ can be written as
\bear
&&{\cal H}=\frac{2}{\eta}\sum_{i=1}^{L-2} \sum_{\alpha=1}^{3} \sigma_{i}^{\alpha}\sigma_{i+1}^{\alpha} + \frac{1}{2\eta}\left( L-1+\frac{1}{(S+\frac{1}{2})^{2}} \right) + \frac{2}{\eta}\left(\frac{1}{S+\frac{1}{2}} \right)^{2}  \sum_{\substack{ \alpha,\beta=1\\ \alpha \neq \beta}}^{3} \sigma_{L-1}^{\alpha} \left\{ S_{L}^{\alpha}, S_{L}^{\beta} \right\} \sigma_{L+1}^{\beta}   \nonumber \\
&&\frac{2}{\eta}\left(\frac{1}{S+\frac{1}{2}} \right)^{2} \sum_{\alpha=1}^{3} \left[  \sigma_{L-1}^{\alpha} S_{L}^{\alpha} + S_{L}^{\alpha} \sigma_{L+1}^{\alpha}  +  2 \sigma_{L-1}^{\alpha} (S_{L}^{\alpha})^{2} \sigma_{L+1}^{\alpha} +\left( \frac{1}{4}-S(S+1) \right) \sigma_{L-1}^{\alpha} \sigma_{L+1}^{\alpha} \right],
\label{hamilt}
\ear
where the index $\alpha=1,2,3$ means the angular momenta components $x,y,z$, respectively.  In
the expression (\ref{hamilt}) is also assumed the following toroidal boundary condition
between the sites $1$ and $L+1$ \cite{RMG}
\EQ
\label{bound}
\left(\begin{array}{c}
        \sigma_{L+1}^{+} \\
        \sigma_{L+1}^{-} \\
        \sigma_{L+1}^{z}
        \end{array}\right)=
        \frac{1}{g_{11}g_{22}-g_{12}g_{21}}\left( \begin{array}{ccc}
        g_{11}^{2} & -g_{21}^{2} & - g_{11}g_{21} \\
        -g_{12}^{2} & g_{22}^{2} & g_{12}g_{22} \\
        -2g_{11}g_{12} & 2g_{21}g_{22} & g_{11}g_{22}+g_{12}g_{21}
                \end{array} \right)\left(\begin{array}{c}
                                \sigma_{1}^{+} \\
                                \sigma_{1}^{-} \\
                                \sigma_{1}^{z}
                                \end{array}\right).
\EN

We see that the 
boundary matrix $\cal G_{A}$ permits a more general type of interactions between the bulk 
spin-$1/2$ and the spin-$S$ impurity. The model (\ref{hamilt},\ref{bound}) turns out to be an interesting extension  of
the impurity system originally proposed in ref.\cite{AJ}.
Finally, the eigenvalues $E^{(\pm)}=\left[\frac{\operatorname{d}}{{\operatorname{d}\lambda}}\ln{\left(\Lambda_{n_{\pm}}^{(\frac{1}{2},\dots,\frac{1}{2},S)}(\lambda) \right)} \right]\vert_{\lambda=0}$ of this Hamiltonian are given by
\EQ
E^{(\pm)}=\frac{1}{\eta} \left[ L-1+\frac{1}{S+\frac{1}{2}} +
\eta^{2}\sum_{i=1}^{n_{\pm}} \frac{1}{[\lambda_{i}^{(\pm)}]^2-\frac{\eta^{2}}{4}} \right],
\EN
where $\lambda_{i}^{(\pm)}$ satisfy the same Bethe ansatz equation (\ref{betheansatz}).

\subsection{The non-compact $SU(1,1)$ realization}\label{noncompactsection}

Another possible realization of the Yang-Baxter algebra (\ref{fundrel}) for the $SU(2)$ $R$-matrix is in terms of the $SU(1,1)$ Lie algebra. One motivation to study integrable models whose quantum space are $SU(1,1)$ invariant comes from the physics of atoms interacting with electromagnetic fields. In fact, different realizations of this symmetry in terms of bosonic operators lead us to a number of solvable atom-fields models \cite{GE,BOG} that are interesting generalizations of  the famous Jaynes and Cummings paradigm \cite{JC}. In this context, the study of these systems with general boundary representations appears to be important, since recently it has been argued \cite{AMI} that suitable combinations between boundary matrices $\cal G_{A}$ and $\cal L$-operators can generated the so-called counter-rotating terms which are relevant in the case of high intensity fields.

The $\cal L$-operator realization in terms of the $SU(1,1)$ generators $K_{j}^{z}$ and $K_{j}^{\pm}$ is given by \cite{SU11},
\EQ
{\cal L}_{{\cal A}j}^{(\frac{1}{2},k_{j})}(\lambda)=\left( \begin{array}{cc}
            \left[ \lambda+c \right] +\eta K_{j}^{z} & -\eta K_{j}^{-} \\
            \eta K_{j}^{+} & \left[ \lambda+c \right] -\eta K_{j}^{z}
            \end{array}\right),
\label{Lsu11}
\EN
where $k_{j}$ is the Bargmann index \cite{BAR} which characterize 
the unitary representations of the $SU(1,1)$ algebra and $c$ is any complex constant. The 
commuting transfer matrix of the corresponding mixed vertex model is therefore,
\EQ
T_{1/2}^{(k_{1},\dots,k_{L})}(\lambda)=\tr_{\cal A}\left[ {\cal G_{A}} {\cal L}_{{\cal A}L}^{(\frac{1}{2},k_{L})} \dots {\cal L}_{{\cal A}1}^{(\frac{1}{2},k_{1})}  \right].
\label{transferT12kj}
\EN

The spectrum of this operator will depend much on the kind of representation we choose for the quantum space. One relevant representation for our purposes is the positive discrete series $D^{+}(k_{j})$ where $k_{j}$ assume integer or half-integer values. More specifically, let the ket $\ket{k_{j},n_{j}}$, with $n_{j}=0,1,2,\dots$ be the basis of $D^{+}(k_{j})$, then the action of the generators is given by
\begin{subequations}
\bear
K_{j}^{z}\ket{k_{j},n_{j}}&=&(n_{j}+k_{j})\ket{k_{j},n_{j}}, \label{propertiesSU11a} \\
K_{j}^{+}\ket{k_{j},n_{j}}&=&\sqrt{(n_{j}+1)(n_{j}+2k_{j})}\ket{k_{j},n_{j}+1}, \\
K_{j}^{-}\ket{k_{j},n_{j}}&=&\sqrt{n_{j}(n_{j}+2k_{j}-1)}\ket{k_{j},n_{j}-1}.
\label{propertiesSU11c}
\ear
\end{subequations}

As before, we should now look for a local reference state $\ket{0}_{j}^{(k_{j})} \in SU(1,1)$ such that the action 
of the transformed 
$M_{j+1}^{-1}{\cal L}_{{\cal A}j}^{(\frac{1}{2},k_{j})}(\lambda)M_j$
on it gives as a result an up triangular matrix.  This condition implies the following annihilation
property
\EQ
\left\{ \left[\lambda+c \right] (p_{j+1}-p_{j}) - \eta K_{j}^{z} (p_{j+1}+p_{j}) 
+\eta K_{j}^{+} p_{j+1} p_{j} + \eta K_{j}^{-} \right\} 
\ket{0}_{j}^{(k_{j})}=0.
\label{Cannihilate1}
\EN

In order to solve  Eq.(\ref{Cannihilate1}) 
we shall repeat the procedure of previous section, i.e. we
expand the pseudovacuum 
$\ket{0}_{j}^{(k_{j})} = \displaystyle \sum_{n_j=0}^{\infty} \bar{C}(k_j,n_j) \ket{k_j,n_j} $
in the $SU(1,1)$ basis and 
use the properties (\ref{propertiesSU11a}-\ref{propertiesSU11c}) as well as that 
$p_j$ is the same for all
$j$ to 
generate relations for the coefficients $\bar{C}(k_j,n_j)$. 
These relations are disentangled recursively
and we find 
that the appropriate state $\ket{0}_{j}^{(k_{j})}$ is 
\EQ
\ket{0}_{j}^{(k_{j})}=\sum_{n_{j}=0}^{\infty} p_{j}^{n_{j}} \sqrt{\frac{(2k_{j}+n_{j}-1)!}{(2k_{j}-1)!n_{j}!}} 
{\bar{C}}(k_{j},0) \ket{k_{j},n_{j}},
\EN
and that the
action of the $\cal L$-operator on it is given by
\EQ
\widetilde{\cal L}_{{\cal A}j}^{(\frac{1}{2},k_{j})}(\lambda) \ket{0}_{j}^{(k_{j})}= \left( \begin{array}{cc}
                         \left( \lambda+c -\eta k_{j} \right)\frac{y_{j}}{y_{j+1}}\ket{0}_{j}^{(k_{j})}  &  \#  \\
                         0   &  \left( \lambda+c +\eta k_{j} \right)\frac{y_{j+1}}{y_{j}}\frac{\det{(M_{j})}}{\det{(M_{j+1})}}\ket{0}_{j}^{(k_{j})}
                        \end{array}\right).
\EN

Before proceeding it is important to remark that the norm of 
$\ket{0}_{j}^{(k_{j})}$ is $|{\bar{C}}(k_j,0)|^2 (1-p_j^2)^{-2k_j}$ which means that this state can be
normalized except when $p_j=\pm 1$. This provides an extra requirement to the entries of the boundary matrix
due to the constraint (\ref{ppm}). Apart from this fact the 
next steps to determine the eigenvalues and eigenfunctions of 
$T_{1/2}^{(k_{1},\dots,k_L)}(\lambda)$ by the quantum 
inverse scattering method are fairly parallel to those 
already described in section \ref{higherspinsection}. Omitting these details the final result for the eigenvalues is
\begin{eqnarray}
\Lambda_{n_{\pm}}^{(k_{1},\dots,k_L)} & = &g_{1/2}^{(\pm)} \prod_{j=1}^{L} \left( \lambda+c-\eta k_{j} \right) \prod_{i=1}^{n_{\pm}}\frac{\lambda_{i}^{(\pm)}-\lambda+\eta-c}{\lambda_{i}^{(\pm)}-\lambda -c} 
\nonumber \\
&&+g_{1/2}^{(\mp)} \prod_{j=1}^{L} \left( \lambda+c+\eta k_{j} \right) \prod_{i=1}^{n_{\pm}}\frac{\lambda-\lambda_{i}^{(\pm)}+\eta+c}{\lambda-\lambda_{i}^{(\pm)}+c},
\end{eqnarray}
provided that the rapidities satisfy the Bethe ansatz equations
\EQ
\prod_{j=1}^{L} \left(\frac{\lambda_{i}^{(\pm)}-\eta k_{j}}{\lambda_{i}^{(\pm)}+\eta k_{j}}\right)=\frac{g_{1/2}^{(\mp)}}{g_{1/2}^{(\pm)}}
\prod_{\stackrel{l=1}{l \neq i }}^{n_{\pm}} \frac{\lambda_{i}^{(\pm)}-\lambda_{l}^{(\pm)}+\eta }{\lambda_{i}^{(\pm)}-\lambda_{l}^{(\pm)}-\eta},
\EN
where we have performed the convenient shift $\lambda_{i}^{(\pm)} \rightarrow \lambda_{i}^{(\pm)}-c$.

We would like to finished this section with the following comment. Having at hand two different families of $\cal L$-operators $\displaystyle {\cal L}_{{\cal A}j}^{(k)}(\lambda) \in \prod_{j=1}^{L_{k}} \otimes V_{j}^{(k)}$ $k=1,2$ which satisfy the Yang-Baxter algebra with the same $R$-matrix, the co-multiplication structure of this algebra allows us to construct even more generalized mixed vertex models by combining the tensor product of these two possible realizations. The corresponding monodromy matrix can be written as
\EQ
{\cal T_{A}}(\lambda)= {\cal G}_{\cal A} \overline{\cal L}_{{\cal A} L_{1}+L_{2}}(\lambda) \overline{\cal L}_{{\cal A} L_{1}+L_{2}-1}(\lambda) \dots \overline{\cal L}_{{\cal A}2}(\lambda) \overline{\cal L}_{{\cal A}1}(\lambda),
\label{monodromyoverline}
\EN
where $\overline{\cal L}_{{\cal A}j}(\lambda)$ is defined by
\EQ
\overline{\cal L}_{{\cal A}j}(\lambda)= \begin{cases}
{\cal L}_{{\cal A}j}^{(1)}(\lambda), ~~ \text{if } j \in \{\gamma_{1}, \dots, \gamma_{L_{1}}\} \\
{\cal L}_{{\cal A}j}^{(2)}(\lambda), ~~ \text{otherwise},
\end{cases}
\EN
and the partition $\{\gamma_{1},\dots,\gamma_{L} \}$ 
denotes a set of integer indices assuming values in the interval $1\leq \gamma_{i} \leq L_{1}+L_{2}$.

For example, one choose as the first family of $\cal L$-operators 
the higher spin operators (\ref{Lhigherspin}) and as the second 
one the $SU(1,1)$ realization (\ref{Lsu11}). Clearly, the eigenvalues problem 
associated with the general monodromy matrix (\ref{monodromyoverline}) 
is solvable by trivial combination of the results of section \ref{su2}. The global 
reference state $\ket{0}^{(\pm)}$ turns  out to be\footnote{ We recall that here we are supposing that the boundary
elements respect the condition $p_j \neq \pm 1$ to make the state $\ket{0}_{j}^{(k_j)}$ 
normalizable.}

\EQ
\ket{0}^{(\pm)}=\prod_{j \in \{\beta_{1},\dots,\beta_{L} \}} \otimes \ket{0}_{j}^{(S_{j})} \prod_{\stackrel{j}{\text{otherwise}}} \otimes \ket{0}_{j}^{(k_{j})},
\EN
while the eigenvalues are
\bear
\Lambda_{n_{\pm}}^{(\{S_j\},\{k_j\})}(\lambda)=g_{1/2}^{(\pm)} \prod_{j=1}^{L_{1}} \left(\lambda +\frac{\eta}{2}+\eta S_{j} \right)\prod_{j=1}^{L_{2}} \left( \lambda+c-\eta k_{j} \right) \prod_{i=1}^{n_{\pm}}\frac{\lambda_{i}^{(\pm)}-\lambda+\eta}{\lambda_{i}^{(\pm)}-\lambda} \nonumber \\
+ g_{1/2}^{(\mp)} \prod_{j=1}^{L_{1}} \left(\lambda +\frac{\eta}{2}-\eta S_{j} \right) \prod_{j=1}^{L_{2}} \left( \lambda+c+\eta k_{j} \right) \prod_{i=1}^{n_{\pm}}\frac{\lambda-\lambda_{i}^{(\pm)}+\eta}{\lambda-\lambda_{i}^{(\pm)}},
\ear
and $\lambda_{i}^{\pm}$ satisfies the Bethe ansatz equation
\EQ
\prod_{j=1}^{L_{1}} \left(\frac{\lambda_{i}^{(\pm)}+\frac{\eta}{2}+\eta S_{j}}{\lambda_{i}^{(\pm)}+\frac{\eta}{2}-\eta S_{j}}\right) \prod_{j=1}^{L_{2}} \left(\frac{\lambda_{i}^{(\pm)}+c-\eta k_{j}}{\lambda_{i}^{(\pm)}+c+\eta k_{j}}\right)=\frac{g_{1/2}^{(\mp)}}{g_{1/2}^{(\pm)}}
\prod_{\stackrel{l=1}{l \neq i }}^{n_{\pm}} \frac{\lambda_{i}^{(\pm)}-\lambda_{l}^{(\pm)}+\eta }{\lambda_{i}^{(\pm)}-\lambda_{l}^{(\pm)}-\eta}.
\EN

\section{$SU(N)$ mixed vertex model}\label{SUNsection}

A natural extension of previous section would be to consider mixed vertex models whose Boltzmann weights that combines isomorphic and non-isomorphic $\cal L$-operators based on the $SU(N)$ symmetry at different orders of representations. Previous results for this mixed model \cite{ALA}, however, suggest us that the details entering the solution of such systems with general twists will follow closely that already described in section \ref{higherspinsection} and in our previous work \cite{RMG}.

However, because $SU(N)$ is not self-conjugate for $N\geq 3$ 
one expects that the Yang-Baxter algebra (\ref{fundrel}) admits further 
classes of realization on this group other than the higher spin 
representations. The purpose of this section is to explore such 
possibility by considering $\cal L$-operators whose quantum 
space is invariant by the conjugate representation of $SU(N)$. Originally 
this $\cal L$-operator was discovered in the context 
of factorizable theories, representing the scattering matrix 
between particles and anti-particles \cite{KUR}. Recently \cite{MAR}, 
this amplitude has been discussed more generally in terms of the braid-monoid algebra. We 
also recall that mixed vertex model combining the fundamental 
and conjugate representations are of direct interest 
in statistical mechanics since it appears to be related 
with the combinatorial problem of coloring the edges 
of the square lattice and fully packed loop models \cite{BER}. 
The  commuting transfer 
matrix $T^{(L_1,L_2)}(\lambda)$
of 
such mixed vertex model with a boundary $\cal G_{A}$ 
is defined by Eqs.(\ref{transfermatrix},\ref{monodromyoverline}). The   
first ${\cal L}_{{\cal A}j}^{(1)}(\lambda)$ operator is the fundamental $SU(N)$ realization obtained
from Eqs.(\ref{VisoA},\ref{R-matrix}), namely 
\EQ
{\cal L}_{{\cal A}j}^{(1)}(\lambda)=\lambda \sum_{\alpha=1}^{N} \hat{e}_{\alpha \alpha}^{(a)} \otimes \hat{e}_{\alpha \alpha}^{(b)} + \eta \sum_{\alpha,\beta=1}^{N} \hat{e}_{\alpha \beta}^{(a)} \otimes \hat{e}_{\beta \alpha}^{(b)},
\label{L-matrix}
\EN
while the second one
${\cal L}_{{\cal A}j}^{(2)}(\lambda)$ 
intertwines between the fundamental and conjugate representation of $SU(N)$ and its expression is \cite{MAR}
\EQ
{\cal L}_{{\cal A}j}^{(2)}(\lambda)=\left(\frac{\lambda}{\eta}-\rho \right) \sum_{\alpha=1}^{N} \hat{e}_{\alpha, \alpha}^{(a)} \otimes \hat{e}_{\alpha, \alpha}^{(b)} - \sum_{\alpha,\beta=1}^{N} \hat{e}_{\alpha, \beta}^{(a)} \otimes \hat{e}_{N+1-\alpha, N+1-\beta}^{(b)}.
\label{L1-matrix}
\EN
where $\rho$ is an extra free parameter.

Our first task is to find the reference states for each transformed operator 
$\widetilde{{\cal L}}_{{\cal A} j}^{(i)}(\lambda)= M_{j+1}^{-1}
{{\cal L}}_{{\cal A} j}^{(i)}(\lambda) M_{j}$ that
annihilates all the $\frac{N(N-1)}{2}$ lower left elements for arbitrary $\lambda$.
The form of such pseudovacuums has a direct dependence on the elements of the gauge matrix which
here will be denoted by 
$\displaystyle M_{j}=\sum_{\alpha,\beta=1}^{N} m_{j}(\alpha,\beta)\hat{e}_{\alpha\beta}^{({\cal A})}$. 
The first local reference state $\ket{0}_j^{(1)}$ has been already determined in ref.\cite{RMG} and 
it is given by the first column of the gauge matrix $M_j$, namely
\EQ
\ket{0}_{j}^{(1)}=\left(\begin{array}{c}
                m_{j}(1,1) \\
                m_{j}(2,1)  \\
                \vdots \\
                m_{j}(N-1,1) \\
                m_{j}(N,1)
                \end{array}\right)_j,
\label{VEC1}
\EN
provided that the following ratios are satisfied
\EQ
p_{\alpha,\beta}=\frac{m_{j}(\alpha,\beta)}{m_{j}(N,\beta)}=
\frac{m_{j+1}(\alpha,\beta)}{m_{j+1}(N,\beta)},~~\mathrm{for}~~ 
\alpha,\beta=1,\dots, N-1.
\label{ppre}
\EN

The second reference state $\ket{0}_j^{(2)}$ has a more involved representation in terms of
the elements of $M_j$. It turns out that this reference state can be put in the following form
\EQ
\ket{0}_{j}^{(2)}=\left(\begin{array}{c}
                c_{N,N}^{(j)} \\
                c_{N-1,N}^{(j)}  \\
                \vdots \\
                c_{2,N}^{(j)} \\
                c_{1,N}^{(j)}
                \end{array}\right)_j,
\label{VEC}
\EN
where $c_{\alpha,\beta}^{(j)}$ are the cofactors of the gauge matrix $M_j$. These are simply obtained from the 
$(\alpha,\beta)$ minors of $M_j$ 
with an appropriate sign, 
\EQ
c_{\alpha,\beta}^{(j)}=(-1)^{\alpha+\beta}\left|\begin{array}{cccccc}
                m_j(1,1) & \dots & m_j(1,\beta-1) & m_j(1,\beta+1) & \dots & m_j(1,N) \\
                \vdots & \ddots      &   \vdots  &    \vdots &       &  \vdots \\
                m_j(\alpha-1,1) & \dots & m_j(\alpha-1,\beta-1) & m_j(\alpha-1,\beta+1) & \dots & m_j(\alpha-1,N) \\
                m_j(\alpha+1,1) & \dots & m_j(\alpha+1,\beta-1) & m_j(\alpha+1,\beta+1) & \dots & m_j(\alpha+1,N) \\
                \vdots &       &   \vdots  &    \vdots & \ddots      &  \vdots \\
                m_j(N,1) & \dots & m_j(N,\beta-1) & m_j(N,\beta+1) & \dots & m_j(N,N) \\
                \end{array}\right|.
\EN

It follows that the
action of the transformed ${\cal L}$-operator 
for fundamental and conjugate representation on their respective reference state is given in terms of
an up
triangular matrix, 
\EQ
\widetilde{{\cal L}}_{{\cal A} j}^{(i)}(\lambda)\ket{0}_{j}^{(i)}=
            \left(\begin{array}{ccccc}
                        a_{1}^{(i)}(\lambda)\frac{f_{1}^{j}}{f_{1}^{j+1}}  & \# & \cdots  & \# & \# \\
                        0 & a_{2}^{(i)}(\lambda)\frac{f_{2}^{j}}{f_{2}^{j+1}}  & \cdots & \# & \# \\
                        \vdots & \vdots & \ddots & \vdots & \vdots \\
                        0 & 0 & \cdots & a_{N-1}^{(i)}(\lambda) \frac{f_{N-1}^{j}}{f_{N-1}^{j+1}}
                        & \# \\
                        0 & 0 & \cdots & 0 & a_{N}^{(i)}(\lambda) \frac{f_{N}^{j}}{f_{N}^{j+1}}
                        \end{array}\right)_{N \times N}\ket{0}_{j}^{(i)},
\label{laxNtriang}
\EN
where $f_{\alpha}^{j}$ are given by
\EQ
f_{\alpha}^{j}= \begin{cases}
m_{j}(N,\alpha), ~~ \alpha=1,\dots ,N-1 \\
\displaystyle \left(\prod_{i=1}^{N-1} \frac{1}{m_{j}(N,i)}\right)
\det{\left(M_{j}\right)}, ~~ \alpha=N,
\end{cases}
\EN
while the rational functions $a_{k}^{(i)}(\lambda)$ are 
\begin{equation}
a_{k}^{(1)}(\lambda)= \begin{cases}
\lambda+\eta~~\mathrm{for}~~k=1  \\
\lambda~~\mathrm{for}~~k=2,\dots,N \end{cases}~~~
a_{k}^{(2)}(\lambda)= \begin{cases}
\frac{\lambda}{\eta}-\rho~~\mathrm{for}~~k=1,\dots,N-1 \\ 
\frac{\lambda}{\eta}-\rho-1~~\mathrm{for}~~k=N. \end{cases}
\end{equation}

The next step is to use the 
the remaining freedom of the elements of the gauge matrix to transform $M_{1}^{-1}{\cal G_{A}}M_{L+1}$ 
into a diagonal matrix.  This condition together with Eq.(\ref{ppre}) impose severe restrictions on the
possible values for the ratios $p_{\alpha,\beta}$ which turns out to be the same satisfied by
the ratio of the components of the eigenvectors of the boundary matrix $\cal G_{A}$. 
This results in $N$ possible choices for $p_{\alpha,1}^{(l)}$, $l=1,\dots,N$ and 
therefore  from Eqs.(\ref{VEC1},\ref{VEC}) 
there exists $N$ type of appropriate local references states that will be
denoted by $\ket{0}_{j}^{(i,l)}$.
As a consequence of that we have $N$ possible choices for the global pseudovacuum which can be written as
\EQ
\ket{0}^{(l)}=\prod_{j \in \{\beta_{1},\dots,\beta_{L} \}} \otimes \ket{0}_{j}^{(1,l)} \prod_{\stackrel{j}{\text{otherwise}}} \otimes \ket{0}_{j}^{(2,l)} ~~~ \text{for} ~~ l=1,\dots,N.
\label{refgen}
\EN

Further progresses are made by choosing a suitable representation for the gauge transformed monodromy matrix 
that are able to distinguish the creation and annihilation fields over the reference state (\ref{refgen}). 
Previous experience with vertex models having the triangular property  (\ref{laxNtriang}) suggest us to take
the structure  used in the nested Bethe ansatz formulation of periodic $SU(N)$ models \cite{SUN,SUN1} 
\EQ
\widetilde{{\cal T_{A}}}(\lambda)=\left(\begin{array}{cccc}
                \widetilde{A}(\lambda) & \widetilde{B}_{1}(\lambda) & \cdots & \widetilde{B}_{N-1}(\lambda) \\
                \widetilde{C}_{1}(\lambda) & \widetilde{D}_{11}(\lambda) & \cdots & \widetilde{D}_{1N-1}(\lambda) \\
                \vdots & \vdots & \ddots & \vdots \\
                \widetilde{C}_{N-1}(\lambda) & \widetilde{D}_{N-11}(\lambda) & \cdots & \widetilde{D}_{N-1 N-1}(\lambda) \\
                \end{array}\right)_{N \times N},
\label{represmonodromy}
\EN

By comparing Eq.(\ref{laxNtriang}) with Eq.(\ref{represmonodromy}) we observe that the operators
$\widetilde{B}_{k}(\lambda)$  produce new states when applied to the pseudovacuum state $\ket{0}^{(l)}$  while
$\widetilde{C}_{k}(\lambda)$ are clearly annihilation fields. Furthermore, we can also read the action of
the diagonal
fields,
\bear
\widetilde{A}(\lambda)\ket{0}^{(l)} & = & [a_{1}^{(1)}(\lambda)]^{L_{1}} [a_{1}^{(2)}(\lambda)]^{L_{2}} \frac{f_{1}^{1}}{f_{1}^{L+1}} \ket{0}^{(l)}, \\
\widetilde{D}_{kk}(\lambda)\ket{0}^{(l)} & =& [a_{k+1}^{(1)}(\lambda)]^{L_{1}} [a_{k+1}^{(2)}(\lambda)]^{L_{2}}  \frac{f_{k+1}^{1}}{f_{k+1}^{L+1}}\ket{0}^{(l)}, ~~~k=1,\dots,N-1.
\ear

We now seek for further  eigenstates $\ket{\phi}^{(l)}$ of $T^{(L_1,L_2)}(\lambda)$  with the following structure
\EQ
\ket{\phi}^{(l)}=\widetilde{B}_{a_{1}}(\lambda_{1}^{(1,l)})\dots \widetilde{B}_{a_{m_{1}^{(l)}}}(\lambda_{m_{1}^{(l)}}^{(1,l)}) {\cal F}^{a_{m_{1}^{(l)}} \dots a_{1}} \ket{0}^{(l)},
\label{MULTI}
\EN
where the indices $a_{j}$ of the coefficients 
${\cal F}^{a_{m_{1}^{(l)}} \dots a_{1}}$  run over $N-1$ possible values. The rapidities $\lambda_j^{(1,l)}$ will be
determined by solving the eigenvalue equation
\EQ
T^{(L_1,L_2)}(\lambda) \ket{\phi}^{(l)}= \Lambda^{(L_1,L_2)}(\lambda) \ket{\phi}^{(l)}.
\EN

At this point we have the basic ingredients  to follow the general strategy of the nested Bethe ansatz method.
We apply either 
$\widetilde{A}(\lambda)$ or
$\widetilde{D}_{kk}(\lambda)$ on $\ket{\phi}^{(l)}$ with the help of commutation rules that are the same known
for the periodic case as explained in our previous work \cite{RMG}. From now on the computations are standard
and since the few adaptations needed have already been described in ref.\cite{RMG} we shall present here
only the final results. We find that the eigenvalues 
$\Lambda_{m_1^{(l)}~\dots~m_{N-1}^{(l)}}^{(L_1,L_2)}(\lambda)$ are given by
\bear
&&\Lambda_{m_1^{(l)}~\dots~m_{N-1}^{(l)}}^{(L_1,L_2)}(\lambda;\{\lambda_{i}^{(1,l)}\}, \dots, \{\lambda_{i}^{(N-1, l)}\})=  g^{(l)} \left( \lambda+\eta\right)^{L_{1}} \left(\frac{\lambda}{\eta}-\rho \right)^{L_{2}} \prod_{j=1}^{m_{1}^{(l)}} \frac{\lambda_{j}^{(1, l)}-\lambda+\eta}{\lambda_{j}^{(1, l)}-\lambda} \nonumber \\
&& +\left(\lambda\right)^{L_{1}} \left( \frac{\lambda}{\eta}-\rho \right)^{L_{2}} \sum_{k=1}^{N-2} g^{(l+k)} \prod_{j=1}^{m_{k}^{(l)}}\frac{ \lambda-\lambda_{j}^{(k,l)} +\eta}{ \lambda-\lambda_{j}^{(k, l)}} \prod_{j=1}^{m_{k+1}^{(l)}} \frac{ \lambda_{j}^{(k+1, l)}-\lambda +\eta}{ \lambda_{j}^{(k+1,l)}-\lambda} \nonumber \\
&& +\left(\lambda \right)^{L_{1}} \left( \frac{\lambda}{\eta}-\rho-1 \right)^{L_{2}} g^{(l+N-1)} \prod_{j=1}^{m_{N-1}^{(l)}} \frac{ \lambda-\lambda_{j}^{(N-1,l)} +\eta}{\lambda-\lambda_{j}^{(N-1, l)}},
%\nonumber \\
\ear
where the phase factors $g^{(l)}$ are the eigenvalues of the boundary gauge matrix $\cal G_{A}$ ordered in
such way that they satisfy the relation 
$g^{(l)}=g^{(l+N)}$ for $l=1,\dots,N$. The corresponding nested Bethe ansatz equations can be written as
\EQ
\frac{g^{(l)}}{g^{(l+1)}} \left[\frac{ \lambda_{i}^{(1,l)} +\eta}{\lambda_{i}^{(1,l)}}\right]^{L_{1}} = \prod_{\stackrel{j=1}{j \neq i}}^{m_{1}^{(l)}} -\frac{ \lambda_{i}^{(1, l)}-\lambda_{j}^{(1,l)} +\eta}{ \lambda_{j}^{(1,l)}-\lambda_{i}^{(1, l)}+\eta} \prod_{j=1}^{m_{2}^{(l)}} \frac{ \lambda_{j}^{(2,l)}-\lambda_{i}^{(1, l)} +\eta}{ \lambda_{j}^{(2, l)}-\lambda_{i}^{(1,l)}},
\label{nestedBA1}
\EN
\bear
\frac{g^{(l+k-1)}}{g^{(l+k)}} \prod_{j=1}^{m_{k-1}^{(l)}}\frac{ \lambda_{i}^{(k,l)}-\lambda_{j}^{(k-1, l)}+\eta}{\lambda_{i}^{(k,l)}-\lambda_{j}^{(k-1, l)}}= \prod_{\stackrel{j=1}{j \neq i}}^{m_{k}^{(l)}}-\frac{ \lambda_{i}^{(k,l)}-\lambda_{j}^{(k,l)}+\eta}{ \lambda_{j}^{(k, l)}-\lambda_{i}^{(k, l)}+\eta}
\prod_{j=1}^{m_{k+1}^{(l)}}\frac{ \lambda_{j}^{(k+1, l)}-\lambda_{i}^{(k, l)}+\eta}{\lambda_{j}^{(k+1, l)}-\lambda_{i}^{(k,l)} }, \\
 k=2, \dots, N-2 \nonumber
\ear
\EQ
\frac{g^{(l+N-2)}}{g^{(l+N-1)}}
\left[\frac{\lambda_{i}^{(N-1,l)}-\eta \rho }{ \lambda_{i}^{(N-1,l)}-\eta (\rho-1)}
\right]^{L_{2}} \prod_{j=1}^{m_{N-2}^{(l)}}
\frac{ \lambda_{i}^{(N-1, l)}-\lambda_{j}^{(N-2,
l)}+\eta}{ \lambda_{i}^{(N-1, l)}-\lambda_{j}^{(N-2, l)}} =
\prod_{\stackrel{j=1}{j \neq i}}^{m_{N-1}^{(l)}}
-\frac{ \lambda_{i}^{(N-1, l)}-\lambda_{j}^{(N-1,
l)}+\eta}{ \lambda_{j}^{(N-1, l)}-\lambda_{i}^{(N-1, l)}+\eta}.
\label{nestedBA2}
\EN

\section{Quantum space transformations}\label{QSTsection}

One striking result of previous sections is that the final forms of the transfer matrix eigenvalues and the Bethe ansatz equations are similar to that expected for vertex models with diagonal boundaries. The corresponding diagonal twists are just the eigenvalues of the non-diagonal boundary matrix $\cal G_{A}$. This result motive us to look for possible correspondences between the eigenvectors of such vertex models with non-diagonal and diagonal boundaries as well.

In order to establish the above mentioned relationship one has to explore the possibility of making quantum space transformations on the gauge transformed $\widetilde{\cal L}$-operators. Rather remarkably, for all the vertex models  discussed in this paper it is always possible to choose a invertible transformations $U_{j}$ on the space $V_{j}$ such that \EQ
U_{j}^{-1} \widetilde{\cal L}_{{\cal A}j}(\lambda) U_{j}={\cal L}_{{\cal A}j}(\lambda).
\label{quantumtransf}
\EN

This means that quantum space transformations can be used to undo the modifications carried out on the $\cal L$-operators by the gauge transformations $M_{j}$. Let us now exemplify the importance of this property on the diagonalization of the standard transfer matrix defined by Eqs.(\ref{transfermatrix},\ref{generalmonodromy}). Denoting 
by $M_{\cal A}$ the matrix that diagonalize the boundary matrix $\cal G_{A}$ we can write $T(\lambda)$ as
\bear
T(\lambda)&=&\tr_{\cal A}{\left[ M_{\cal A} D_{\cal A} M_{\cal A}^{-1} M_{\cal A} \left( M_{\cal A}^{-1} {\cal L}_{{\cal A}L}(\lambda) M_{\cal A} \right) \dots \left( M_{\cal A}^{-1} {\cal L}_{{\cal A}1}(\lambda) M_{\cal A} \right) M_{\cal A}^{-1} \right]}, \label{tttransfer} \\
&=&\tr_{\cal A}{\left[ D_{\cal A} \widetilde{\cal L}_{{\cal A}L}(\lambda) \widetilde{\cal L}_{{\cal A}L-1}(\lambda) \dots \widetilde{\cal L}_{{\cal A}1}(\lambda) \right]},
\label{nada}
\ear
where $D_{\cal A}$ is diagonal matrix whose entries are the eigenvalues of $\cal G_{A}$ and the $\widetilde{\cal L}$-operators are given by
\EQ
\widetilde{\cal L}_{{\cal A}j}(\lambda)=M_{\cal A}^{-1} {\cal L}_{{\cal A}j}(\lambda)M_{\cal A}.
\label{auxiliarytransf}
\EN

Now motivated by 
property (\ref{quantumtransf}) we can define a new operator $T'(\lambda)$
\bear
T'(\lambda)=\prod_{j=1}^{L} \otimes U_{j}^{-1} T(\lambda) \prod_{j=1}^{L} \otimes U_{j} 
=\tr_{\cal A}{\left[  D_{\cal A} {\cal L}_{{\cal A}L}(\lambda)  \dots {\cal L}_{{\cal A}1}(\lambda) \right]},
\label{TlinhaT}
\ear
which is precisely the transfer matrix of the vertex model we have started with with diagonal twist $D_{\cal A}$.

At this point is important to recall that the transfer matrix $T'(\lambda)$ can be diagonalized with very little difference from the periodic case because the diagonal boundary $D_{\cal A}$ does not change drastically the properties of the monodromy matrix elements. This not only explain the reason why the Bethe ansatz equations and eigenvalues have the same shape for both diagonal and non-diagonal boundaries but also makes it possible to substantiate a clear relation between their eigenvectors. In fact, if we denote by $\ket{\psi'}$ one possible eigenstate of $T'(\lambda)$ then it follows directly from (\ref{TlinhaT}) that the corresponding eigenkets  $\ket{\psi}$ of $T(\lambda)$ will be given by
\EQ
\ket{\psi}=\prod_{j=1}^{L} \otimes U_{j} \ket{\psi},
\label{rel}
\EN
and that the eigenvalues of $T'(\lambda)$ and $T(\lambda)$ are of course identical.

Explicit expressions for the quantum matrices $U_{j}$ are summarized in appendix B for 
all the vertex models discussed  so far. It turns out that the above observations can be used to diagonalize the transfer matrix of even more complicated mixed vertex models. For instance, let us consider the following operator
\EQ
T_{g}(\lambda)=\tr{\left[ {\cal G_{A}}^{(L)} {\cal L}_{{\cal A}L}(\lambda) {\cal G_{A}}^{(L-1)} {\cal L}_{{\cal A}L-1}(\lambda) \dots {\cal G_{A}}^{(1)} {\cal L}_{{\cal A}1}(\lambda) \right]},
\label{manyboundaries}
\EN
that combines boundary matrices and $\cal L$-operators at any site of the lattice. 

Assuming 
that ${\cal G_{A}}^{(j)}$ are non-singular matrices one 
can insert ${\cal G_{A}}^{(j)}\left[{\cal G_{A}}^{(j)} \right]^{-1} $ terms 
all over the trace (\ref{manyboundaries}), permitting  us to rewrite the transfer matrix $T_{g}(\lambda)$ as
\EQ
T_{g}(\lambda)=\tr{\left[ {\cal G_{A}}^{(ef)} \widetilde{\cal L}_{{\cal A}L}(\lambda) \widetilde{\cal L}_{{\cal A}L-1}(\lambda) \dots \widetilde{\cal L}_{{\cal A}1}(\lambda) \right]},
\label{TRAN}
\EN
where the new boundary matrix ${\cal G_{A}}^{(ef)}$ is given in terms of the following ordered product
\EQ
{\cal G_{A}}^{(ef)}={\cal G_{A}}^{(L)} {\cal G_{A}}^{(L-1)} \dots {\cal G_{A}}^{(1)}.
\label{ordered}
\EN

The transformed $\widetilde{\cal L}$-operators in Eq.(\ref{TRAN}) are 
given by an extension of formula (\ref{auxiliarytransf}) that accommodates site dependent transformations, namely
\EQ
\widetilde{\cal L}_{{\cal A}j}(\lambda)=\left[ M_{\cal A}^{(j)} \right]^{-1} {\cal L}_{{\cal A}j}(\lambda)M_{\cal A}^{(j)}, \EN
in such way that for each site $j$ the matrices $M_{\cal A}^{(j)}$ are
\EQ
M_{\cal A}^{(j)}={\cal G_{A}}^{(j-1)} {\cal G_{A}}^{(j-2)} \dots {\cal G_{A}}^{(1)}.
\EN

As before, the next step is to find the appropriate quantum transformations $U_{j}$  in order to reverse the
gauge transformations $M_{\cal A}^{(j)}$ in  Eq.(\ref{TRAN}). 
Note that the matrix $U_j$ are now different for each site $j$ because 
its elements depend crucially on the entries of $M_{\cal A}^{(j)}$, see appendix B for details. By performing the
transformation (\ref{TlinhaT}) we are able 
to turn the transfer matrix
$T_{g}(\lambda)$ into
\EQ
T'_{g}(\lambda)=\tr_{\cal A}{\left[ {\cal G_{A}}^{(ef)} {\cal L}_{{\cal A}L}(\lambda) {\cal L}_{{\cal A}L-1}(\lambda) \dots {\cal L}_{{\cal A}1}(\lambda) \right] }.
\EN

Once again we end up with a transfer matrix problem 
with only  one boundary matrix that has been 
discussed in details through this paper. Interesting enough, the 
diagonal twists entering into the Bethe ansatz equations and the transfer matrix eigenvalues are exactly 
the eigenvalues of the ordered product (\ref{ordered}) of all the boundary matrices. We remark that our
findings when applicable to the two-site transfer 
matrix $T_{1/2}^{(S_1,S_2)}(\lambda)$ of section 2.1 
reproduce the eigenvalues and the 
Bethe ansatz equations proposed recently
in ref.\cite{AMI} except for the fact that 
the number of roots vary up to $2(S_1 +S_2)$ instead of being fixed at this
upper bound value. 

In what follows we will present two concrete examples of the utility 
of quantum space transformations in the problem of transfer matrices diagonalization.

\subsection{Spin-$S$ Heisenberg model}\label{spinSSsection}

The classical analogue of the integrable spin-$S$ Heisenberg model 
\cite{KUL,BATA} is known to be a $2S+1$ state vertex model 
whose Boltzmann weights are identified with the matrix elements 
of the following $SU(2)$ invariant $\cal L$-operator \cite{KUL,BATA}
\EQ
{\cal L}_{{\cal A}j}^{(S)}(\lambda)= \sum_{l=0}^{2S}
f_{l}(\lambda) P_{l},
\EN
where $\displaystyle f_{l}(\lambda)=(\lambda+2 \eta S)\prod_{k=l+1}^{2S} \frac{\lambda- \eta k}{\lambda+ \eta k}$ 
and $P_{l}$ is the projector onto $SU(2)_{l}$ in the 
Clebsch-Gordan decomposition $SU(2)_{S} \otimes SU(2)_{S}$. This operator is conveniently represented by the expression
\EQ
P_{l}=\prod_{\stackrel{k=0}{k
\neq l}}^{2S} \frac{\vec{S}\otimes \vec{S}_{j}-x_{k}}{x_{l}-x_{k}},
\EN
with $x_{k}=\frac{1}{2}\left[l(l+1)-2S(S+1) \right]$.

As usual, the transfer matrix associated to this $SU(2)$ vertex model with an integrable boundary ${\cal G_{A}}$ is
\EQ
T_{S}(\lambda)=\tr_{\cal A}{\left[ {\cal G_{A}} {\cal L}_{{\cal A}L}^{(S)}(\lambda) \dots {\cal L}_{{\cal A}1}^{(S)}(\lambda) \right]}.
\label{transferSS}
\EN

After some computation we find that the most general boundary condition compatible with integrability can be written as
\EQ
{\cal G_{A}}=\left(\begin{array}{cccc}
                g_{S, S} & g_{S, S-1} & \dots & g_{S, -S} \\
                g_{S-1, S} & g_{S-1, S-1} & \dots & g_{S-1, -S} \\
                \vdots & \vdots & \ddots & \vdots \\
                g_{-S, S} & g_{-S, S-1} & \dots & g_{-S, -S} \\
                \end{array}\right),
\label{gssmatrix}
\EN
where the coefficients $g_{m,l}$ satisfy the recurrence relation
\EQ
g_{m, l}=\frac{\sqrt{2S(S-m)(S+1+m)} g_{S-1, S}
g_{m+1, l} - \sqrt{2S(S-l)(S+1+l)} g_{S, S-1} g_{m, l+1}}{2S(l-m)
g_{S, S}},
\EN
for $l \neq m$ and $l,m=-S,\dots, S$ and
\EQ
g_{l,l}=\frac{g_{S-1, S-1} g_{l+1, l+1}}{g_{S, S}}-\frac{
\sqrt{2(2S-1)(S-l-1)(S+l+2)} g_{S-2, S} g_{l+2, l} + 2(S-l-1)
g_{S-1, S} g_{l+1,l}}{\sqrt{2S(S+l+1)(S-l)} g_{S, S}},
\EN
for $l=-S,\dots, S$.

As expected from the $SU(2)$ symmetry this boundary matrix has four free parameters represented by the elements $g_{S,S}, g_{S,S-1}, g_{S-1,S}$ and $g_{S-1,S-1}$. The standard procedure used to diagonalize transfer matrix based on higher spin
representations consists in exploring its commutation 
with that of the mixed vertex  model (\ref{transferT12S}) when 
$S_{1}=S_{2}=\dots=S_{L}=S$. Unfortunately, this property is no 
longer valid for arbitrary non-diagonal boundaries and therefore 
another route has to be taken in these cases. One possible way is to 
proceed exactly as explained in the beginning of section \ref{QSTsection}. We first 
find the matrix $M_{\cal A}^{(S)}$ that diagonalize the boundary matrix $\cal G_{A}$ (\ref{gssmatrix}) and
afterwards we perform the quantum space transformation (\ref{quantumtransf}) by using the 
matrices $U_{j}^{(S)}$ collected in the appendix B. The transformed 
transfer matrix $T'_{S}(\lambda)$ is then given by
\EQ
T'_{S}(\lambda)=\tr_{\cal A}{\left[ D_{\cal A}^{S} {\cal L}_{{\cal A}L}^{(S)}(\lambda) 
\dots {\cal L}_{{\cal A}1}^{(S)}(\lambda)  \right]},
\label{transferlinha}
\EN
where $D_{\cal A}^{S}$ is the diagonal matrix having the following two possible forms
\EQ
{D_{\cal A}^{S}}_{(+)}=
\displaystyle \diag \left(g_{S}^{(+)}, g_{S}^{(-)}, 
\frac{({g_{S}^{(-)})}^{2}}{g_{S}^{(+)}}, \dots, 
\frac{{(g_{S}^{(-)})}^{S-m}}{{(g_{S}^{(+)})}^{S-m-1}}, \dots, 
\frac{{(g_{S}^{(-)})}^{2S-1}}{{(g_{S}^{(+)})}^{2S-2}}, 
\frac{{(g_{S}^{(-)})}^{2S}}{{(g_{S}^{(+)})}^{2S-1}}  \right),
\EN
and
\EQ
{D_{\cal A}^{S}}_{(-)}=
\displaystyle \diag \left( \frac{{(g_{S}^{(-)})}^{2S}}{{(g_{S}^{(+)})}^{2S-1}}, \frac{{(g_{S}^{(-)})}^{2S-1}}{{(g_{S}^{(+)})}^{2S-2}}, \dots, \frac{{(g_{S}^{(-)})}^{S+m}}{{(g_{S}^{(+)})}^{S+m-1}}, 
\dots, \frac{({g_{S}^{(-)})}^{2}}{g_{S}^{(+)}}, g_{S}^{(-)}, g_{S}^{(+)} \right).
\label{DSpmmatrix}
\EN

The parameters 
$g_{S}^{(\pm)}$ are two particular eigenvalues that are able to parameterize the total
eigenspectrum of the boundary matrix (\ref{gssmatrix}). We recall that these two possibilities we have at our
disposal is related to a remaining ${\cal{Z}}_2$ symmetry of the Hilbert space.
Now $T'_{S}(\lambda)$ can be diagonalized by the quantum inverse scattering method adapting the results 
of ref.\cite{BATA} to include diagonal twists. It turns out that the 
final result for the eigenvalues of $T_{S}(\lambda)$ is
\EQ
\Lambda_{n_{\pm}}^{(S)}(\lambda)=\sum_{m=-S}^{S} \frac{(g_{S}^{(-)})^{S \mp m}}{(g_{S}^{(+)})^{S \mp m-1}} 
\left[ t_{m}(\lambda) \right]^{L} \prod_{j=1}^{n_{\pm}} q_{m}(\lambda-\lambda_{j}^{(\pm)}+\frac{\eta}{2}),
\EN
where the functions $t_{m}(\lambda)$ and $q_{m}(\lambda)$ are given by
\EQ
t_{m}(\lambda)=(\lambda+2 \eta S)\prod_{k=m+1}^{S} \frac{\lambda+ \eta k -\eta S}{\lambda+\eta k + \eta S},
\label{alphamSS}
\EN
and
\EQ
q_{m}(\lambda)=\frac{(\lambda+\eta S+\frac{\eta}{2})(\lambda - \eta S-\frac{\eta}{2})}{(\lambda+ \eta m+\frac{\eta}{2})(\lambda+ \eta m-\frac{\eta}{2})},
\EN
while the corresponding 
Bethe ansatz equation for the rapidities $\lambda_j^{(\pm)}$ are
\EQ
 \left[\frac{\lambda_{i}^{(\pm)}+\eta S_{j}}{\lambda_{i}^{(\pm)}-\eta S_{j}}\right]^{L}= 
\frac{g_{S}^{(\mp)}}{g_{S}^{(\pm)}}
\prod_{\stackrel{l=1}{l \neq i }}^{n_{\pm}} 
\frac{\lambda_{i}^{(\pm)}-\lambda_{l}^{(\pm)}+\eta }{\lambda_{i}^{(\pm)}-\lambda_{l}^{(\pm)}-\eta}.
\label{BAnsatzS}
\EN

\subsection{Singular boundary matrix}

In the most part of this paper 
we have implicitly supposed that the boundary matrices $\cal G_{A} $ are invertible. Here we would
like to explore the interesting situation in which these matrices become singular. For sake of
simplicity we shall discuss this problem for the transfer matrix 
$T_{1/2}^{(\frac{1}{2},\dots,\frac{1}{2})}(\lambda)$ (\ref{transferT12S})
that are built up in terms of the simplest 
$\cal{L}$-operators ${\cal{L}}_{{\cal{A}}j}^{(\frac{1}{2},\frac{1}{2})}(\lambda)$.
A possible
representation of the corresponding singular boundary matrix is 
\EQ
{\cal G}_{\cal A}^{(sg)} = \left( \begin{array}{cc}
        g_{11} & \frac{g_{11}g_{22}}{g_{21}} \\
        g_{21} & g_{22}
        \end{array} \right),
\EN
where we are assuming that $g_{21} \neq 0$.

For general values of $g_{ij}$ this matrix is still diagonalizable and it is similar to a diagonal
matrix ${\cal{D}}^{(sg)}$, namely 
\EQ
{\cal{D}}^{(sg)} = \left( \begin{array}{cc}
        0 & 0 \\
        0 & g_{11}+g_{22}
        \end{array} \right),
\EN
while the explicit expression for the matrix $M_{\cal A}$  that diagonalize 
${\cal G}_{\cal A}^{(sg)}$ is 
\EQ
M_{\cal A}^{(sg)} = \left( \begin{array}{cc}
        -\frac{g_{22}}{g_{21}} & \frac{g_{11}}{g_{21}} \\
        1 & 1
        \end{array} \right).
\label{sing}
\EN

We now proceed by performing the gauge transformations (\ref{nada}) 
with the help of the matrix (\ref{sing}) as well as the quantum space transformation (\ref{quantumtransf})
where
$U_j=M_{\cal A}^{(sg)}$. As a result, the transformed transfer matrix we need to diagonalize becomes
\EQ
{T'}_{1/2}^{(\frac{1}{2},\dots,\frac{1}{2})}(\lambda)=\tr_{\cal A}{\left[D_{\cal A}^{(sg)} 
{\cal L}_{{\cal A} L}^{(\frac{1}{2},\frac{1}{2})}(\lambda)\dots 
{\cal L}_{{\cal A} 1}^{(\frac{1}{2},\frac{1}{2})}(\lambda) \right]}.
\EN

Clearly, this transfer matrix is defective since 
${\cal{D}}^{(sg)}$  contains one diagonal vanishing element and therefore it is proportional to a single 
diagonal monodromy
matrix element. Thanks to this property, the non-null eigenstates $\ket{\psi'}^{(n)}$ of this operator can be found by
direct inspection and they are given by
\EQ
\ket{\psi'}^{(n)}=\prod_{i=1}^{n} \otimes \begin{pmatrix} 1 \\ 0  
\end{pmatrix}_{i} 
\prod_{i=n+1}^{L} \otimes \begin{pmatrix} 0 \\ 1 \end{pmatrix}_{i},
\label{vectorTlinha}
\EN
for $n=0,1,\dots,L$.  The corresponding eigenvalues 
$\Lambda^{(n)}(\lambda)$ have the following form
\EQ
\Lambda^{(n)}(\lambda)= (g_{11}+g_{22}) [\lambda]^{n} [\lambda+\eta]^{(L-n)}.
\label{ggg}
\EN

Considering the result (\ref{vectorTlinha}) and Eq.(\ref{rel}) with 
$U_j=M_{\cal A}^{(sg)}$ we conclude that the original eigenstates of 
$T_{1/2}^{(\frac{1}{2},\dots,\frac{1}{2})}(\lambda)$
are
\EQ
\ket{\psi}^{(n)}=\prod_{i=1}^{n} \otimes \begin{pmatrix} -\frac{g_{22}}{g_{21}} \\ 1  
\end{pmatrix}_{i} \prod_{i=n+1}^{L} \otimes \begin{pmatrix} \frac{g_{11}}{g_{21}} \\ 1 \end{pmatrix}_{i}.
\label{eee}
\EN

Finally, we remark that the results (\ref{ggg},\ref{eee}) have been previously conjectured in ref.\cite{RMG}
on basis of exact diagonalization of $T_{1/2}^{(\frac{1}{2},\dots,\frac{1}{2})}(\lambda)$ 
for some values of $L$. This discussion
not only clarifies the origin of such results 
but also emphasizes the importance of the quantum space 
transformations in the general framework
of transfer matrices diagonalization.

\section{Conclusions}\label{Conclusion}

In this paper we have investigated integrable non-diagonal toroidal boundary conditions for 
mixed vertex models and related spin chains based on the isotropic $SU(N)$ $R$-matrix. In particular,
we have applied the quantum inverse scattering method to obtain the transfer matrix eigenvalues 
and the corresponding Bethe ansatz equations for systems with quantum spaces 
invariant by the high spin 
representation of $SU(2)$, the discrete $D^{+}(k)$ representation of $SU(1,1)$ and the conjugate
representation of $SU(N)$.

We have introduced the notion of quantum 
space transformations that together with the standard 
Baxter's gauge transformations \cite{BA} 
allowed us to diagonalize the
transfer matrix of general families of vertex models that mix
both boundary and $\cal L$-operator representations. We have used this approach to solve the
Heisenberg model with arbitrary spin and non-diagonal twists as well as the case of singular
boundary matrices.  We strong believe that this method is easily applied 
to  other isotropic vertex models such as those invariant 
by the $O(N)$ and $Sp(2N)$ Lie algebras \cite{RE}. We also hope 
that this property could be 
generalized to accommodate the solution of 
trigonometric vertex models and therefore improving
our understanding of the integrability.

\section*{Acknowledgements}
The author G.A.P. Ribeiro thank  FAPESP (Funda\c c\~ao de Amparo \`a Pesquisa do Estado de S\~ao Paulo) 
for financial support. The work of M.J. Martins has been support by the Brazilian Research Council-CNPq and FAPESP.

\addcontentsline{toc}{section}{Appendix A}
\section*{\bf Appendix A: Completeness for $SU(2)$ mixed vertex models}
\setcounter{equation}{0}
\renewcommand{\theequation}{A.\arabic{equation}}

In this appendix we consider the 
completeness of the Bethe ansatz solution of the $SU(2)$ invariant vertex models discussed in section
\ref{higherspinsection} for $L=2$.  More precisely, we are going 
to verify that all the
$(2S_1+1)(2S_2+1)$ eigenvalues of the transfer matrix 
$T_{1/2}^{(S_1,S_2)}(\lambda)$ (\ref{transferT12S}) can be obtained
from the Eqs.(\ref{eingmeioS},\ref{betheansatz}) for some values of $S_1$ and $S_2$:

\begin{itemize}
\item $T_{1/2}^{(\frac{1}{2},1)}(\lambda)$
\end{itemize}

We can search for solutions of the 
Bethe ansatz equations (\ref{betheansatz}) beginning either with
$\ket{0}^{(+)}$ or 
$\ket{0}^{(-)}$ eigenvectors. By choosing the state  
$\ket{0}^{(+)}$  the zero-particle eigenvalue can directly be read from (\ref{eingmeioS}),
\EQ
\Lambda_{0}^{(\frac{1}{2},1)}(\lambda)=g_{1/2}^{(+)}(\lambda+\eta)(\lambda+\frac{3\eta}{2}) + 
g_{1/2}^{(-)}(\lambda)(\lambda-\frac{\eta}{2}).
\EN

For the one-particle state 
$\widetilde{B}(\lambda_{1}^{(+)})\ket{0}^{(+)}$ we find two possible Bethe ansatz roots given by
\EQ
\lambda_{1\pm}^{(+)}=\frac{ 3\eta \left( g_{1/2}^{(-)} + g_{1/2}^{(+)} \right) \pm 
\eta \sqrt{(g_{1/2}^{(-)}+g_{1/2}^{(+)})^2 + 32 g_{1/2}^{(-)} g_{1/2}^{(+)} }}{4\,\left( g_{1/2}^{(-)} - g_{1/2}^{(+)} \right) },
\EN
and the respective eigenvalues are
\EQ
\Lambda_{1 \pm}^{(\frac{1}{2},1)}(\lambda)=g_{1/2}^{(+)} \left[ \lambda^2+\frac{\eta}{2}\lambda +\frac{\eta^2}{4} \right] +g_{1/2}^{(-)} \left[ \lambda^2+\frac{\eta}{2}\lambda -\frac{\eta^2}{4} \right] \pm \sqrt{(g_{1/2}^{(-)}+g_{1/2}^{(+)})^{2}+32 g_{1/2}^{(+)}g_{1/2}^{(-)}}.
\EN

The next step is to consider the two-particle state
$\widetilde{B}(\lambda_{1}^{(+)})\widetilde{B}(\lambda_{2}^{(+)})\ket{0}^{(+)}$ and again two possibilities
are obtained and they are
\bear
\lambda_{1\pm}^{(+)}=\frac{5\eta \left( g_{1/2}^{(-)} + g_{1/2}^{(+)} \right)}
{8 \left( g_{1/2}^{(-)} - g_{1/2}^{(+)} \right)}  \mp \frac{\eta\sqrt{(g_{1/2}^{(-)}+g_{1/2}^{(+)})^2 
+ 32 g_{1/2}^{(-)} g_{1/2}^{(+)}}}{8 \left( g_{1/2}^{(-)} - g_{1/2}^{(+)} \right)} \nonumber \\
+\frac{\eta \sqrt{2} \sqrt{5 ({g_{1/2}^{(-)}}^2 - {g_{1/2}^{(+)}}^2) - 38 g_{1/2}^{(-)} g_{1/2}^{(+)} \pm 3\left( g_{1/2}^{(-)} + g_{1/2}^{(+)} \right)  \sqrt{(g_{1/2}^{(-)}+g_{1/2}^{(+)})^2 + 32 g_{1/2}^{(-)} g_{1/2}^{(+)}}}}{8 \left( g_{1/2}^{(-)} - g_{1/2}^{(+)} \right)},
\ear
and 
\bear
\lambda_{2\pm}^{(+)}=\frac{5\eta \left( g_{1/2}^{(-)} + g_{1/2}^{(+)} \right)}
{8 \left( g_{1/2}^{(-)} - g_{1/2}^{(+)} \right)}  \mp \frac{\eta\sqrt{(g_{1/2}^{(-)}+g_{1/2}^{(+)})^2 
+ 32 g_{1/2}^{(-)} g_{1/2}^{(+)}}}{8 \left( g_{1/2}^{(-)} - g_{1/2}^{(+)} \right)} \nonumber \\
-\frac{\eta \sqrt{2} \sqrt{5 ({g_{1/2}^{(-)}}^2 - {g_{1/2}^{(+)}}^2) - 38 g_{1/2}^{(-)} g_{1/2}^{(+)} \pm 3\left( g_{1/2}^{(-)} + g_{1/2}^{(+)} \right)  \sqrt{(g_{1/2}^{(-)}+g_{1/2}^{(+)})^2 + 32 g_{1/2}^{(-)} g_{1/2}^{(+)}}}}{8 \left( g_{1/2}^{(-)} - g_{1/2}^{(+)} \right)}.
\ear

The corresponding eigenvalues are then given by
\EQ
\Lambda_{2\pm}^{(\frac{1}{2},1)}(\lambda)=g_{1/2}^{(-)} \left[ \lambda^2+\frac{\eta}{2}\lambda +\frac{\eta^2}{4} \right] +g_{1/2}^{(+)} \left[ \lambda^2+\frac{\eta}{2}\lambda -\frac{\eta^2}{4} \right] 
\pm \sqrt{(g_{1/2}^{(-)}+g_{1/2}^{(+)})^{2}+32 g_{1/2}^{(+)}g_{1/2}^{(-)}}.
\EN

The last sector consists of three Bethe ansatz roots  and an analytical expression for them 
has been prevented within  our numerical resources. We have however performed an extensive numerical work 
confirming that the associated eigenvalue is
\EQ
\Lambda_{3}^{(\frac{1}{2},1)}(\lambda)=g_{1/2}^{(-)}(\lambda+\eta)(\lambda+\frac{3\eta}{2}) + 
g_{1/2}^{(+)}(\lambda)(\lambda-\frac{\eta}{2}).
\EN

\begin{itemize}
\item $T_{1/2}^{(1,1)}(\lambda)$
\end{itemize}

As before, starting with the state $\ket{0}^{(+)}$ it is not difficult to see that
\EQ
\Lambda_{0}^{(1,1)}(\lambda)=g_{1/2}^{(+)} [\lambda+\frac{3\eta}{2}]^2+g_{1/2}^{(-)}[\lambda-\frac{\eta}{2}]^2.
\EN

The next state is the one-particle sector giving us two possibilities
\EQ
\lambda_{1\pm}^{(+)}=-\eta\left( \frac{\sqrt{g_{1/2}^{(+)}} \pm \sqrt{g_{1/2}^{(-)}}}{\sqrt{g_{1/2}^{(+)}} \mp \sqrt{g_{1/2}^{(-)}}} \right),
\EN
and the corresponding eigenvalues are 
\EQ
\Lambda_{1\pm}^{(1,1)}(\lambda)=g_{1/2}^{(+)} (\lambda +\frac{3\eta}{2})(\lambda+\frac{\eta}{2})
+g_{1/2}^{(-)} (\lambda-\frac{\eta}{2})(\lambda+\frac{\eta}{2})
\pm \sqrt{g_{1/2}^{(+)} g_{1/2}^{(-)}}.
\EN

On the other hand the two-particle state provides us three types of Bethe ansatz roots. The first
one is
\EQ
\lambda_{1}^{(+)}=\eta\frac{(g_{1/2}^{(-)}+g_{1/2}^{(+)})}{(g_{1/2}^{(-)}-g_{1/2}^{(+)})}~~\mathrm{and}~~
\lambda_{2}^{(+)}=0,
\EN
and the respective eigenvalue is 
\EQ
\Lambda_{2}^{(1,1)}(\lambda)= \left( g_{1/2}^{(+)} + g_{1/2}^{(-)}\right) 
(\lambda+\frac{3\eta}{2})(\lambda-\frac{\eta}{2}).
\EN

The other remaining solutions are given by
\bear
\lambda_{1\pm}^{(+)}=\frac{3\eta(g_{1/2}^{(-)}+g_{1/2}^{(+)})}{4(g_{1/2}^{(-)}-g_{1/2}^{(+)})} 
\mp \eta \frac{\sqrt{(g_{1/2}^{(+)}+g_{1/2}^{(-)})^2+
32 g_{1/2}^{(+)}g_{1/2}^{(-)}}}{4(g_{1/2}^{(-)}-g_{1/2}^{(+)})} \nonumber \\
+\eta\frac{ \sqrt{(g_{1/2}^{(+)}+g_{1/2}^{(-)})^2-16 g_{1/2}^{(+)}g_{1/2}^{(-)} \pm ({g_{1/2}^{(-)}}^2-{g_{1/2}^{(+)}}^{2})\sqrt{(g_{1/2}^{(+)}+g_{1/2}^{(-)})^2+32 g_{1/2}^{(+)}g_{1/2}^{(-)}}}}{2\sqrt{2}(g_{1/2}^{(-)}-g_{1/2}^{(+)})},
\ear
and 
\bear
\lambda_{2\pm}^{(+)}=\frac{3\eta(g_{1/2}^{(-)}+g_{1/2}^{(+)})}{4(g_{1/2}^{(-)}-g_{1/2}^{(+)})} 
\mp \eta \frac{\sqrt{(g_{1/2}^{(+)}+g_{1/2}^{(-)})^2+
32 g_{1/2}^{(+)}g_{1/2}^{(-)}}}{4(g_{1/2}^{(-)}-g_{1/2}^{(+)})} \nonumber \\
-\eta\frac{ \sqrt{(g_{1/2}^{(+)}+g_{1/2}^{(-)})^2-16 g_{1/2}^{(+)}g_{1/2}^{(-)} \pm ({g_{1/2}^{(-)}}^2-{g_{1/2}^{(+)}}^{2})\sqrt{(g_{1/2}^{(+)}+g_{1/2}^{(-)})^2+32 g_{1/2}^{(+)}g_{1/2}^{(-)}}}}{2\sqrt{2}(g_{1/2}^{(-)}-g_{1/2}^{(+)})},
\ear
while the associated eigenvalues are
\EQ
\Lambda_{2 \pm}^{(1,1)}(\lambda)=\left( g_{1/2}^{(+)} + g_{1/2}^{(-)}\right) 
\left[ (\lambda+ \frac{3\eta}{2})(\lambda-\frac{\eta}{2}) 
+\frac{\eta}{2} \right] \pm \frac{1}{2} \sqrt{(g_{1/2}^{(+)}+g_{1/2}^{(-)})^{2}+32g_{1/2}^{(+)}g_{1/2}^{(-)}}.
\EN

The remaining sectors have been investigated numerically and we verified that the three and the four
particle states produce the following eigenvalues
\EQ
\Lambda_{3\pm}^{(1,1)}(\lambda)=g_{1/2}^{(-)} (\lambda +\frac{3\eta}{2})(\lambda+\frac{\eta}{2})
+g_{1/2}^{(+)} (\lambda-\frac{\eta}{2})(\lambda+\frac{\eta}{2})
\pm \sqrt{g_{1/2}^{(+)} g_{1/2}^{(-)}},
\EN
and
\EQ
\Lambda_{4}^{(1,1)}(\lambda)=g_{1/2}^{(-)} [\lambda+\frac{3\eta}{2}]^2+g_{1/2}^{(+)}[\lambda-\frac{\eta}{2}]^2.
\EN

Finally, we remark that as byproduct of this analysis we have also checked the completeness of the
eigenspectrum of the transfer matrix $T_1^{(1,1)}(\lambda)$ of section \ref{spinSSsection} since its Bethe
ansatz equation are the same as that discussed above.

\addcontentsline{toc}{section}{Appendix B}
\section*{\bf Appendix B: Quantum space transformation}
\setcounter{equation}{0}
\renewcommand{\theequation}{B.\arabic{equation}}

In this appendix we will present the explicit expressions for the quantum space transformations $U_j$ 
that are able to undo the gauge transformations on the auxiliary space. The matrix elements of $U_j$ are
directly related to entries of the gauge matrix $M_{\cal A}^{(j)}$ since they have been required
to satisfy the following identity
\EQ
{\cal{L}}_{{\cal{A}}j}(\lambda)= U_j^{-1}[M_{\cal A}^{(j)}]^{-1} 
{\cal{L}}_{{\cal{A}}j}(\lambda) M_{\cal A}^{(j)} U_j,
\EN
for each $\cal L$-operator mentioned in the main text.
In order to present this relationship it is convenient to write the gauge matrix as
\EQ
M_{\cal A}^{(j)}=\sum_{\alpha,\beta=1}^{N} \bar{m}_{\alpha,\beta}^{(j)} \hat{e}_{\alpha,\beta}^{(\cal A)},
\EN
where we recall that $N$ is the dimension of the auxiliary space $\cal A$.
We now start to list the quantum matrices $U_j$ for the $\cal L$-operators used in this paper 

\begin{itemize}
\item For ${\cal{L}}_{{\cal A}j}^{(\frac{1}{2},S_j)}(\lambda)$:
\end{itemize}
\EQ
U_j^{(\frac{1}{2}, S_j)}= \left(\begin{array}{cccc}
                        u_{S_j,S_j} & u_{S_j,S_j-1} & \cdots & u_{S_j,-S_j} \\
                        u_{S_j-1,S_j} & u_{S_j-1,S_j-1} & \cdots & u_{S_j-1,-S_j} \\
                        \vdots & \vdots & \ddots & \vdots \\
                        u_{-S_j,S_j} & u_{-S_j,S_j-1} & \cdots & u_{-S_j,-S_j} \\
                        \end{array}\right),
\EN
where some of the matrix elements $u_{i,j}$ are 
\EQ
u_{S_j,S_j-k} = [\bar{m}_{1,1}^{(j)}]^{2S_j-k}[\bar{m}_{1,2}^{(j)}]^{k} \sqrt{\frac{(2S_j)!}{k!(2S_j-k)!}}, 
\EN
while the remaining ones satisfy the following recurrence relation
\bear
u_{S_j-k-1,S_j-n}=\frac{(\bar{m}_{1,1}^{(j)} \bar{m}_{2,2}^{(j)}- \bar{m}_{1,2}^{(j)} \bar{m}_{2,1}^{(j)})}{[\bar{m}_{1,1}^{(j)}]^{2} } 
\sqrt{ \frac{n(2S_j-n+1)}{(k+1)(2S_j-k)}} u_{S_j-k,S_j-n+1} \nonumber \\
+\left( \frac{\bar{m}_{2,1}^{(j)}}{\bar{m}_{1,1}^{(j)}} \right)^{2}
\sqrt{\frac{k(2S_j-k+1)}{(k+1)(2S_j-k)}}u_{S_j-k+1,S_j-n} +\frac{\bar{m}_{2,1}^{(j)}}{\bar{m}_{1,1}^{(j)}} \frac{2(S_j-k)}{\sqrt{(k+1)(2S_j-k)}} u_{S_j-k,S_j-n},
\ear
and $k,n$ are integers satisfying $ k,n=0, \dots, 2S_j$.

\begin{itemize}
\item For ${\cal{L}}_{{\cal A}j}^{(\frac{1}{2},k_j)}(\lambda)$:
\end{itemize}
\EQ
U_j^{(\frac{1}{2}, k_j)}= \left(\begin{array}{cccc}
                        \bar{u}_{0,0} & \bar{u}_{0,1} &  \bar{u}_{0,2} & \cdots  \\
                        \bar{u}_{1,0} & \bar{u}_{1,1} &  \bar{u}_{1,2} & \cdots  \\
			\bar{u}_{2,0} & \bar{u}_{2,1} &  \bar{u}_{2,2} & \cdots  \\
			\vdots & \vdots & \vdots & \ddots \\
                        \end{array}\right),
\label{Ukquantum}
\EN
where the first row are given by
\EQ
\bar{u}_{0,n} = (-1)^{n} \left( \frac{\bar{m}_{2,1}^{(j)}}{\bar{m}_{2,2}^{(j)}} 
\right)^{n} \sqrt{\frac{(2k_j+n-1)!}{n!(2k_j-1)!}} \bar{u}_{0,0}  
\EN
and the other elements are obtained recursively by the following relation
\bear
\bar{u}_{n+1,l}=\frac{(\bar{m}_{1,1}^{(j)} \bar{m}_{2,2}^{(j)}- \bar{m}_{1,2}^{(j)} \bar{m}_{2,1}^{(j)})}{[\bar{m}_{2,2}^{(j)}]^{2}} 
\sqrt{ \frac{l(2k_j+l-1)}{(n+1)(2k_j+n)}} \bar{u}_{n,l-1} \nonumber \\
- \left(\frac{\bar{m}_{1,2}^{(j)}}{\bar{m}_{2,2}^{(j)}} \right)^{2} \sqrt{\frac{n(2k_j+n-1)}{(n+1)(2k_j+n)}}  
\bar{u}_{n-1,l} +\frac{\bar{m}_{1,2}^{(j)}}{\bar{m}_{2,2}^{(j)}} \frac{2(k_j+n)}{\sqrt{(n+1)(2k_j+n)}} \bar{u}_{n,l},
\ear
and $n,l$ are integers $n,l=0, 1, 2, \dots$. 

\begin{itemize}
\item For the higher spin operator ${\cal{L}}_{{\cal A}j}^{(S)}(\lambda)$:
\end{itemize}
\EQ
U_j^{(S)}= M_{\cal A}^{(S)}
\EN

\begin{itemize}
\item For the fundamental $SU(N)$ operator ${\cal{L}}_{{\cal A}j}^{(1)}(\lambda)$:
\end{itemize}
\EQ
U_j^{(1)}= M_{\cal A}^{(j)}
\EN

\begin{itemize}
\item For the conjugated $SU(N)$ operator ${\cal{L}}_{{\cal A}j}^{(2)}(\lambda)$:
\end{itemize}
\EQ
U_j^{(2)}=\left(\begin{array}{cccc}
                        \bar{c}_{N,N}^{(j)} & \bar{c}_{N,N-1}^{(j)} & \cdots & \bar{c}_{N,1}^{(j)} \\
                        \bar{c}_{N-1,N}^{(j)} & \bar{c}_{N-1,N-1}^{(j)} & \cdots & \bar{c}_{N-1,1}^{(j)} \\
                        \vdots & \vdots & \ddots & \vdots \\
                        \bar{c}_{1,N}^{(j)} & \bar{c}_{1,N-1}^{(j)} & \cdots & \bar{c}_{1,1}^{(j)} \\
                        \end{array}\right),
\EN
where $\bar{c}^{(j)}_{\alpha,\beta}$ are the following cofactors 
\EQ
\bar{c}_{\alpha,\beta}^{(j)}=(-1)^{\alpha+\beta}\left|\begin{array}{cccccc}
                \bar{m}_{1,1}^{(j)} & \dots & \bar{m}_{1,\beta-1}^{(j)} & \bar{m}_{1,\beta+1}^{(j)} 
& \dots & \bar{m}_{1,N}^{(j)} \\
                \vdots & \ddots      &   \vdots  &    \vdots &       &  \vdots \\
                \bar{m}_{\alpha-1,1}^{(j)} & \dots & \bar{m}_{\alpha-1,\beta-1}^{(j)} 
& \bar{m}_{\alpha-1,\beta+1}^{(j)} & \dots & \bar{m}_{\alpha-1,N}^{(j)} \\
                \bar{m}_{\alpha+1,1}^{(j)} & \dots & \bar{m}_{\alpha+1,\beta-1}^{(j)} 
& \bar{m}_{\alpha+1,\beta+1}^{(j)} & \dots & \bar{m}_{\alpha+1,N}^{(j)} \\
                \vdots &       &   \vdots  &    \vdots & \ddots      &  \vdots \\
                \bar{m}_{N,1}^{(j)} & \dots & \bar{m}_{N,\beta-1}^{(j)} & \bar{m}_{N,\beta+1}^{(j)} 
& \dots & \bar{m}_{N,N}^{(j)} \\
                \end{array}\right|.
\EN


\addcontentsline{toc}{section}{References}
\begin{thebibliography}{}
\bibitem{FA}  L.A. Takhtajan and L.D. Faddeev, {\em Russian Math. Surveys, 34 (1979) 11}
\bibitem{KO} V.E. Korepin, G. Izergin and N.M. Bogoliubov, {\em Quantum 
Inverse Scattering Method and Correlation Functions, Cambridge University Press, 1993}
\bibitem{KUL} P.P. Kulish, N.Y. Reshetikhin and E.K. Sklyanin, {\em Lett.Math.Phys. 5 (1981) 393}
\bibitem{BATA} L.A. Takhtajan, {\em Phys.Lett. A 87 (1982) 479};
H.M. Babujian, {\em Nucl.Phys. B, 215 (1983) 317}
\bibitem{QG1} M. Jimbo, {\em Lett. Math.Phys. 10 (1985) 63};
V.G. Drinfeld, {\em Quantum Groups, Proc.Int.Cong. of Math., 1986, AM Gleason, p. 798}
\bibitem{QG2} N.Y. Reshetikhin, L.A. Takhtajan, L.D. Faddeev, {\em Leningrad.Math.Journal 1 (1990) 193}
\bibitem{DE} H.J. de Vega, {\em Nucl.Phys.B, 240 (1984) 495}
\bibitem{BA1} C.M. Yung and M.T. Bachelor, {\em Nucl.Phys.B, 446 (1995) 461}; M.T. Batchelor, R.J. Baxter,
M.J.O. Rourke and C.M. Yung, {\em J.Phys.A:Math.Gen. 28 (1995) 2759}
\bibitem{SUN} P.P. Kulish and N.Y. Reshetikhin, {\em Sov.Phys.JETP 53 (1981) 108; 
J.Phys.A: Math.Gen. 16 (1983) L591}
\bibitem{SUN1} O. Babelon, H.J. de Vega and C.M. Viallet, {\em Nucl.Phys.B, 200 (1982) 266}
\bibitem{RMG} G.A.P. Ribeiro, M.J. Martins and W. Galleas, {\em Nucl. Phys.B, 675 (2003) 567}
\bibitem{AJ} N. Andrei and H. Johannesson, {\em Phys.Lett.A 100 (1984) 108 }
\bibitem{DW} H.J. de Vega and F. Woyanorovich, {\em J.Phys.A: Math.Gen. 25 (1992) 4499}
\bibitem{BA} R.J. Baxter, ``Exactly Solved Models in Statistical Mechanics'', Academic Press, New York, 1982.
\bibitem{GE} C.C. Gerry, {\em Phys.Rev.A 7 (1988) 2683}; V. Buzek, {\em Phys.Rev.A 39 (1989) 3196}
\bibitem{BOG} N.M. Bogoliubov, R.K. Bullough and J. Timonen, {\em J.Phys.A: Math.Gen. 29 (1996) 6305}; 
A. Rybin, G. Kastelewicz, J. Timonen and N. Bogoliubov, {\em J.Phys.A: Math.Gen. 31 (1998) 4705}
\bibitem{JC} E.T. Jaynes and F.W. Cummings, {\em Proc.IEEE 51 (1963) 89}
\bibitem{AMI} L. Amico and K. Hikami, {\em cond-mat/0309680}
\bibitem{SU11} B. Jurco, {\em J.Math.Phys. 30 (1989) 1739}; 
V.B. Kuznetsov and A.V. Tsiganov, {\em J.Phys.A: Math.Gen. 22 (1989) L73}; 
V.Y. Chernyak, A.E. Kozhekin and E.I. Ogievetsky, {\em J.Phys.A: Math.Gen. 26 (1993) 1313}
\bibitem{BAR} V. Bargmann, {\em Ann.Math., 48 (1947) 568}
\bibitem{ALA} S.R. Aladim and M.J. Martins, {\em J.Phys.A:Math.Gen. 26 (1993) 7287}
\bibitem{KUR} B. Berg, M. Karowski, V. Kurak and P. Weisz, {\em Nucl.Phys.B 134 (1978) 125}
\bibitem{MAR} M.J. Martins, 
``Series on Advances in Statistical Mechanics'', 
Eds. M.T. Batchelor and L.T. Wille, World Scientific, Singapore, 1999, p. 425
\bibitem{BER} B. Nienhuis, {\em J.Stat.Phys. 102 (2001) 981}; D. Dei Cont and B. Nienhuis, {\em J.Phys.A:Math.Gen. 37 (2004) 3085}
\bibitem{RE} N.Yu. Reshetikhin, {\em Theor.Math.Phys. 63 (1985) 555}
\end{thebibliography}
\end{document}